%% file: resubmission2.tex
\documentclass[10pt, 
aps,
prx,
twocolumn,
noeprint,
longbibliography,
nofootinbib
]{revtex4-2}

\input{preamble.tex}

\graphicspath{{figures/}}
\usepackage{color}
\newcommand{\pdag}{{\phantom{\dagger}}}
\newcommand{\bigcdot}{\mathbin{\raisebox{-0.25ex}{\scalebox{1.7}{$\cdot$}}}}

\begin{document}

\title{Towards Efficient Quantum Thermal State Preparation via Local Driving:\\
Lindbladian Simulation with Provable Guarantees}

\author{Dominik Hahn}
\author{S. A. Parameswaran}
\author{Benedikt Placke}
\affiliation{Rudolf Peierls Centre for Theoretical Physics, University of Oxford, Oxford OX1 3PU, United Kingdom}

\begin{abstract}
Preparing the thermal density matrix $\rho_{\beta}\propto e^{-\beta H}$ corresponding to a given Hamiltonian $H$ is a task of central interest across quantum many-body physics, and is particularly salient when attempting to study it with quantum computers. Although solved {in principle} by recent constructions of efficiently simulable Lindblad master equations  --- that provably have $\rho_{\beta}$ as a steady state [C.-F. Chen {\it et al}, {\it Nature} 646, pp. 561–566 (2025)] --- the implementation of these ``exact Gibbs samplers'' requires large-scale quantum computational resources and is hence challenging {in practice} on current or even near-term quantum devices. 
Here, we propose a scheme for approximately
 simulating an exact Gibbs sampler that only requires the [repeated] implementation of three readily available ingredients: (a) analog simulation of $H$; (b) strictly local but time-dependent couplings to ancilla qubits; and (c) reset of the ancillas.
We give rigorous guarantees on the difference between the fixed point reached by our protocol and the exact thermal state, which only depend on parameters of the protocol and its \emph{mixing time}. The procedure is efficiently implementable on near-term devices if $H$ is local, and the mixing time scales mildly with both system size and protocol parameters. While guaranteeing the latter for Hamiltonians of interest remains an important problem for future work,  here we lay the groundwork for developing fully efficient thermal state preparation protocols on quantum simulators.
\end{abstract}

\maketitle

\section{Introduction}

Preparing thermal states of many-body systems is a key goal for a wide range of quantum devices, since it enables the exploration of such systems through ``quantum numerical experiments''. However, the ability of analog quantum simulators to access low-temperature regimes of target systems such as the Hubbard model is often limited by the absence of scalable, problem-agnostic techniques for removing entropy in a controlled manner so as to reach the target energy density. Similarly, while a proposed near-term use-case for digital quantum computers is to attack problems in quantum chemistry, this goal is often obstructed by the absence of efficient ways to prepare a thermal density matrix. While various physically-motivated approaches have been proposed to meet this challenge \cite{lloyd1996universal, terhal2000equilibration,Riera_2012,Ge_2016, Mozgunov2020completelypositive,McArdle_2019, shtanko2021preparing, schuckert2023probing, mi2024quasiparticle, lloyd2024quasiparticle,deshpande2024dynamicparameterizedquantumcircuits,Consiglio_2025}, these are typically of a heuristic and case-by-case nature. As such, their systematic error is poorly understood: in other words, it is often unclear how exactly one has to scale the available resources to obtain a close approximation to a given target state. 

In counterpoint to this are recent rigorous results in the quantum computer science literature \cite{poulin2009sampling, bilgin2010preparing, temme2011quantum_metropolis, rall2023roundin, yung2012quantum_quantum_metropolis, chowdhury2017quantum, motta2020determining, holmes2022quantum, wocjan2023szegedy, zhang2023dissipative, jian2024quantum_metropolis}. A subset of particular interest to this work concerns {\it quantum Gibbs sampling} \cite{kastoryano2016commuting, chen2021fast, CKBG23, CKG23, gilyen2024glauber, ding2024efficient}: the problem of engineering a dissipative quantum dynamics whose steady state corresponds to a specified Gibbs density matrix.  Although a formal solution has long been available in terms of so-called Davies generators \cite{davies1974,davies1976,guo2025designing}, the corresponding dynamics involves a purely dissipative Lindblad evolution under highly nonlocal and hence unphysical `jump operators', making it unfeasible in practice. The recent work upends this conventional wisdom by demonstrating that Gibbs sampling is possible with controllable accuracy (in the sense above) while inducing dissipation using only {\it quasi-local} jump operators, but at the cost of introducing a specially tailored coherent evolution. The steady state of the resulting Lindblad evolution is exactly given by $\rho_\beta$, which follows from the fact that the Lindbladian satisfies a certain [quantum] detailed balance property~\cite{agarwal1973,alicki1976detailed,frigerio1977quantum,fagnola2007generators}. The sole remaining unknown is the mixing time $\tau_{\rm mix}$ of the Lindbladian, which  controls the approach to the steady state and can be long for physically meaningful reasons much as in the classical case. Thus, although  Gibbs samplers are unlikely to speed up problems that are classically hard because of glassy landscapes \cite{Santoro2002Theory,Altshuler2010Anderson,Bapst2013The,rakovszky2024bottleneck,placke2024slow_mixing, garmarnik2024slow_mixing, anschuetz2024glassiness} (such as generic optimization problems), they are not limited by the sign problem or entanglement growth which are the usual ``intrinsically quantum'' obstacles to simulating many-body physics.

\begin{figure}
    \centering
    \includegraphics[width=\columnwidth]{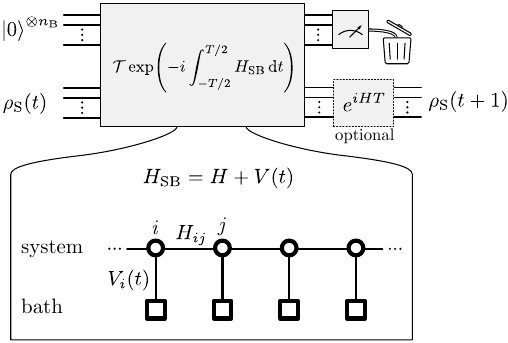}
    \caption{Schematic of one time step of the local driving sampler. In each step, the ``bath'' qubits are initialized in the all-zero state, followed by an entangling step consisting of time evolution with the system Hamiltonian $H$ together with a dynamically driven system-bath coupling $V(t) = \sum_i V_i(t)$ [cf. \autoref{eq:HSB_schroedinger}]. Finally, the bath qubits are measured and the outcomes discarded. The final unitary ``rewinding'' $e^{iHT}$ can be dropped with only a modest increase to the error bound, and can be ignored with no penalty if a quasiparticle picture applies.
    }
    \label{fig:setup}
\end{figure}

Despite this remarkable breakthrough, convenient protocols to implement these exact Gibbs samplers (or their close approximations) in physically realistic settings relevant to current and near-term quantum hardware remain largely unexplored (but see e.g. Refs. \onlinecite{chen2024randomized, brunner2024lindblad_engineering} for concurrent work). Proposed approaches rely on block-encoding Lindbladians \cite{CKG23,ding2024efficient}; doing so at appreciable scale is likely out of the question for the near term. It is this gap between formal theory and practical implementation that we bridge here, while bolstering a physically motivated picture with mathematical rigor.

Specifically, we devise a protocol which \emph{approximately} prepares thermal states of an arbitrary Hamiltonian, with controlled error $\epsilon$, using a total Hamiltonian simulation time scaling as $\Tilde O(\beta \tau_{\rm mix}^3 \epsilon^{-2})$. 
Crucially, we only require three relatively standard features of the current generation of quantum devices:
(a) ``analog'' simulation of the $n_{\rm S}$-qubit Hamiltonian; (b) a dynamically tunable coupling to an (arbitrary) number of $n_{\rm B}$ ancilla qubits [usually, we will take $n_{\rm B}=\order{n_{\rm S}}$]; and (c) the ability to reset the ancillas (see \autoref{fig:setup}).
  
The proposed protocol implements a controlled discrete-time approximation of a certain Lindblad evolution. The corresponding Lindbladian has exactly the same dissipative part as the exact sampler, but the ``wrong'' coherent part: a Lamb shift correction rather than the coherent dynamics necessary for exact thermal state preparation. However, using the so-called secular approximation \cite{breuer_petruccione2002,correa2023lamb_shift,Mozgunov2020completelypositive}, we show that this difference in coherent evolution only incurs a controlled, arbitrary small error per Lindbladian simulation time step. Assuming that the mixing time, and hence the simulation time necessary to reach the steady state, is short, this implies also that the fixed point of our protocol is close to the thermal state. 
In practice, the mixing time generally depends on parameters of the protocol which are also used to control the error per time step. This limits the minimum achievable fixed-point error. Developing a better understanding of , and eventually overcoming, these limitations---which also impede the practicality of the exact samplers---remains an important open question raised by our work.

Aficionados of near-term quantum algorithms will recognize a resemblance between our proposal and the ``quasiparticle cooling'' approach of Refs. \onlinecite{mi2024quasiparticle,lloyd2024quasiparticle}. Formally, the difference between the two {protocols} is that ours implements a certain additional unitary ``rewinding'' procedure. However, this step is unnecessary {\it if} one assumes the validity of a quasiparticle picture. We note that our motivation here is also distinct: we aim to engineer a specific time evolution of the system-bath coupling, such that the dissipative portion of our Lindbladian coincides with that of an exact Gibbs sampler. This then admits an analysis that does not rely on a quasi-particle picture, allowing us to arrive at rigorous error bounds at arbitrary temperature and for arbitrary Hamiltonians; {\it inter alia}, our results also provide a rigorous (but less tight) error bound for the quasiparticle cooling protocol. Nevertheless, the link to Refs.~\onlinecite{mi2024quasiparticle,lloyd2024quasiparticle}  
highlights the feasibility of our protocol. 
It also suggests one reason why their approach to thermal state preparation appears to work even when a quasiparticle description is absent (see Ref.~\onlinecite{lloyd2024quasiparticle} for examples), and that it can be fashioned into a versatile tool with broader applicability.

The remainder of this paper is organized as follows. In \autoref{sec:exactGibbs}, we provide some relevant background on the history of quantum thermal state preparation, with an emphasis on recent work on exact Gibbs samplers. (The latter may also serve as a useful physicists' guide to recent results in the quantum computer science literature.)  We then introduce our local driving protocol in \autoref{sec:LDS}, and both state and motivate bounds on its effectiveness and its efficiency, in terms of the scaling of resources required to achieve a specified proximity (in $1$-norm distance) to the target density matrix. We then proceed to give a rigorous proof of these bounds in \autoref{sec:proofs}, which contains the bulk of the technical material. Readers primarily interested in physical intuition can skip directly to \autoref{sec:resources} that gives the resource estimate implied by these bounds. \autoref{sec:numerics} simulates the protocol numerically on small systems, demonstrating agreement with various aspects of the error bounds. We close with a discussion in \autoref{sec:conclusions}, where we also outline promising directions for future study.

\section{State Preparation, Detailed Balance, and Exact Gibbs Sampling \label{sec:exactGibbs}}

In order to introduce concepts and place our work in context, we first give some relevant background thermal state preparation. After a brief historical orientation, our main emphasis is on recent constructions of Lindblad operators whose steady state corresponds to a specified density matrix $\rho$, by virtue of their satisfying a  so-called Kubo–Martin–Schwinger (KMS) detailed balance condition with respect to $\rho$.

\subsection{Historical Background}
As noted in the introduction, the use of quantum computers to study many body systems in thermal equilibrium usually requires the preparation of a thermal (Gibbs) density matrix $\rho_\beta = Z^{-1} e^{-\beta H}$ (where $Z \equiv \text{tr}\, e^{-\beta H}$) of a given Hamiltonian $H$. A conceptually significant effort to meet this challenge was made by Terhal and DiVincenzo \cite{terhal2000equilibration}, who proposed adapting the physical picture of equilibration under system-bath dynamics to an algorithmic, computational setting. They argued that repeated cycles of initializing a bath of ancilla qubits, evolving its interaction with the system for finite time (e.g. using Hamiltonian simulation) and performing a reset of the ancillas,  asymptotically prepare a thermal state of the system. However, this result is exact only in the limit of infinite bath size and vanishing system bath coupling familiar in the derivation of master equations. Thus, their analysis does not address the feasibility of the protocol in more practical settings, where making general statements in the absence of any foreknowledge of $H$ is challenging.

Ref.~\onlinecite{temme2011quantum_metropolis} instead proposed an algorithm of a very different spirit, by lifting the technique of importance-sampling configurations from the classical to the quantum setting. Such ``quantum Metropolis sampling'' involves a Metropolis-weighted random walk on eigenstates, using quantum phase estimation  to determine energies.

It has recently been claimed that this algorithm is not  provably efficient due to the finite energy resolution: the error in estimating the energy is argued to propagate to the error (e.g. in norm distance) between the actual and desired fixed points of the sampling algorithm, viewed as a quantum channel on the Hilbert space. (For a recent claimed resolution via weak measurements, see Ref.~\onlinecite{jian2024quantum_metropolis}.)

A similar issue also applies to system-bath models. One way of ensuring that the Lindbladian governing the evolution of the system has $\rho_\beta$ as a fixed point is to demand that it satisfy quantum detailed balance\footnote{As in the classical case, this is sufficient  but not necessary.}; while we will make this notion precise below, a lucid discussion oriented towards physicists may be found in Ref.~\onlinecite{guo2025designing}. A formal solution satisfying detailed balance, the so-called ``Davies' Lindbladian'', requires non-local jump operators that distinguish individual eigenstates, so that its construction again requires exponentially fine energy resolution (or equivalently, exponentially long time evolution). Quantum Metropolis sampling relies on access to the eigenbasis to ensure  detailed balance. Thus, both approaches fail due to the well-known~\cite{schrodinger_statistical_thermodynamics} unphysicality of eigenstates  in many-body systems. 

Alternative approaches have employed classical detailed balance together with the eigenstate thermalization hypothesis to construct Gibbs samplers~\cite{shtanko2021preparing,brunner2024lindblad_engineering}. 
However, this assumption may limit their applicability in low-temperature regimes or in systems that fail to act as their own bath.

A recent burst of activity has provided an elegant {\it exact} solution to the longstanding challenge of preparing thermal states. First, Ref.~\onlinecite{CKBG23} observed that for any geometrically local Hamiltonian $H$, the Davies' jump operators can be replaced by `filtered' analogues that are quasilocal (in essence, generated by finite time evolution of local operators) yet leading to a Lindbladian whose fixed point is only polynomially far from the thermal state $\rho_\beta$. Shortly thereafter, a subset of those authors showed~\cite{CKG23} that adding a suitably tailored coherent evolution leads to a Lindbladian with $\rho_\beta$ as an {\it exact} fixed point. The central insight in both works is that one should impose a suitable form of ``Kubo-Martin-Schwinger'' (KMS) detailed balance condition on the Lindbladian, and adjust the filter function and corresponding coherent evolution to ensure that this is satisfied. This construction was then generalized~\cite{gilyen2024glauber,ding2024efficient}, leading to the identification of the most general family of Lindbladians satisfying the KMS condition. We turn next to a summary of these recent results, which will serve as essential background to our work.

\subsection{Exact Gibbs Samplers}

Since KMS detailed balance and exact Gibbs sampling play a central role in our work,  we give a brief overview of these topics to establish notation and to make this paper self-contained. Since we take a somewhat abstract perspective when introducing the necessary formalism, readers may wish to initially skip or just skim this section, and return to it after the more concrete derivations of \autoref{sec:LDS}, but before studying the derivation of error bounds in \autoref{sec:proofs}.

Much of what follows, including our notation, is adapted from  Refs.~\onlinecite{CKG23,ding2024efficient}, to which the reader is referred for details. While we attempt to give as much detail as feasible, more laborious computations are relegated to \appref{app:DBderivation}.

\subsubsection{Kubo-Martin-Schwinger Detailed Balance}

The Lindbladian  superoperator implements time evolution of states (density matrices) in the Schr\"odinger picture, according to \cite{lindblad1976equation}
\begin{equation}
\frac{d\rho}{dt} = \mathcal{L}[\rho],
\end{equation}
so that formally we have $\rho(t) =e^{t\mathcal{L}}[\rho(0)] $. For $\mathcal{L}$ to represent a sensible time evolution, we require that $e^{t\mathcal{L}}$ is a completely positive trace-preserving (CPTP) map on density matrices. Trace preservation $\text{tr}\,e^{t\mathcal{L}}[\rho] =\text{tr}\,\rho$ in turn requires that $\text{tr}\,\mathcal{L}[\rho] =0$.

The corresponding Heisenberg evolution of operators is implemented by the adjoint Lindbladian $\mathcal{L}^\dagger$, taken with respect to the Frobenius inner product $\langle A, B \rangle \equiv \text{tr}[A^\dagger B]$: for any state $\rho$ and operator $\mathcal{O}$, we have $\tr[\mathcal{L}[\rho]\mathcal{O}] = \tr[\rho\mathcal{L}^\dagger[\mathcal{O}]]$. Requiring that the Schr\"odinger evolution preserves the trace is equivalent to requiring that the Heisenberg evolution be {\it unital}, i.e. preserves the identity,  $e^{t\mathcal{L}^\dagger}[\mathds{1}] = \mathds{1}$, which in turn requires $\mathcal{L}^\dagger[\mathds{1}] = 0$. 

Given any full-rank density matrix $\rho$, we  also define a self-adjoint ``weighting'' superoperator $\Gamma_\rho$: 
\begin{equation}
\Gamma_{\rho}[\bigcdot] := \rho^{1/2}(\bigcdot)\rho^{1/2} =\Gamma_\rho^\dagger[\bigcdot] .
\end{equation}

A Lindbladian $\mathcal{L}$ satisifies KMS detailed balance  with respect to a full-rank density matrix $\rho$ if
\begin{equation}
    \mathcal{L}^\dagger =  \Gamma_\rho^{-1}\circ \mathcal{L} \circ \Gamma_\rho.\label{eq:rhoDB}
\end{equation}
An immediate corollary of KMS detailed balance is that $\rho$ is a fixed point of $\mathcal{L}$. To see this, observe that \autoref{eq:rhoDB} is equivalent to the condition 
 $   \mathcal{L} =  \Gamma_\rho\circ \mathcal{L}^\dagger \circ \Gamma_\rho^{-1}$, 
so that we have
\begin{align}
\mathcal{L}[\rho] &= \rho^{1/2} \mathcal{L}^\dagger[ \rho^{-1/2}(\rho)\rho^{-1/2}]\rho^{1/2} 
=0,
\end{align}
where the second equality follows from $\mathcal{L}^\dagger[\mathds{1}]=0$. 

In its stated form [\autoref{eq:rhoDB}], the KMS condition does not immediately resemble the classical notion of detailed balance, which is often formulated in terms of the ``reversibility'' of the transition probabilities. The connection can be sharpened  by defining the {\it KMS inner product} with respect to $\rho$:
\begin{equation}
\langle A, B \rangle_\rho \equiv \text{tr} \sqrt{\rho}A \sqrt{\rho} B.\label{eq:KMS_inner}
\end{equation}
The KMS detailed balance condition [\autoref{eq:rhoDB}] is then the statement that $\mathcal{L}^\dagger$ is self-adjoint with respect to the inner product in \autoref{eq:KMS_inner}. Equivalently, we may also say that $\mathcal{L}\circ \Gamma_\rho =\Gamma_\rho \circ \mathcal{L}^\dagger$ is self-adjoint with respect to the Frobenius inner product. Either perspective brings the KMS condition in consonance with classical detailed balance, which can be defined as the transition matrix being symmetric when weighted appropriately by the Boltzmann weights.

\subsubsection{Structure of KMS-Detailed Balanced Lindbladians}

To understand the structure imposed on $\mathcal{L}$  by KMS detailed balance,  it is convenient to pass to a representation in terms of the Bohr frequencies (i.e. energy differences) of the Hamiltonian $H$ that specifies the target thermal density matrix $\rho_\beta$; we denote the set of such frequencies $\nu\in \bohr$. We may then write $A(t) \equiv e^{iHt} A e^{-iHt} = \sum_{\nu\in \bohr} A_\nu e^{i\nu t}$, where 
\begin{equation}\label{eq:Bohrdef}
    A_\nu \equiv \sum_{\substack{\omega_1, \omega_2 \in {\rm Spec}(H) \\\omega_1-\omega_2=\nu}} \ket{\omega_1}\bra{\omega_1} A\ket{\omega_2}\bra{\omega_2},
\end{equation} 
so that $(A_{-\nu})^\dagger = (A^\dagger)_\nu$.  

We will first write $\mathcal{L}$ in terms of the Bohr frequency representation of some set of jump operators $A_a \in \mathcal{A}$. 
In this representation, a generic Lindbladian $\mathcal{L} = \mathcal{G}+\mathcal{T}+\mathcal{R}$ can be decomposed into ``transition'', ``decay'', and ``coherent parts'' given by 
\begin{subequations}
\begin{align}
\mathcal{T}[\bigcdot] &= \sum_{a\in \mathcal{A}}\sum_{\nu_1, \nu_2\in\bohr} \alpha_{\nu_1, \nu_2} A^a_{\nu_1}(\bigcdot) (A^a_{\nu_2})^\dagger,\label{eq:Tdef}\\
\mathcal{R}[\bigcdot] &=  -\frac{1}{2}\sum_{a\in \mathcal{A}}\sum_{\nu_1, \nu_2} \alpha_{\nu_1, \nu_2}\left\{(A^a_{\nu_2})^\dagger A^a_{\nu_1},\bigcdot\right\},\label{eq:Rdef}\\
\mathcal{G}[\bigcdot] &=  -i\sum_{a\in \mathcal{A}}\sum_{\nu_1, \nu_2} g_{\nu_1, \nu_2}\left[(A^a_{\nu_2})^\dagger A^a_{\nu_1},\bigcdot\right],\label{eq:Gdef}
\end{align}
\label{eq:TRGdefs}    
\end{subequations}

where $\sum_{a\in \mathcal{A}}\ldots \equiv\sum_{a=1}^{|\mathcal{A}|}$, and we require that $\alpha_{\nu_1, \nu_2} = (\alpha_{\nu_2, \nu_1})^*$,  $g_{\nu_1, \nu_2} = (g_{\nu_2, \nu_1})^*$ and that $\alpha_{\nu_1, \nu_2}$ is a positive semidefinite matrix for $\mathcal{L}$ to be CPTP. The  potentially counterintuitive ordering of the frequency labels in $\mathcal{T}$ relative to $\mathcal{R}$ and $\mathcal{G}$ is required by the Lindbladian structure.

Note that while the kernel $\alpha_{\nu_1, \nu_2}$ of $\mathcal{T}$ is fixed to that of $\mathcal{R}$ by trace preservation, that of the coherent part  $g_{\nu_1, \nu_2}$ is arbitrary. However, as we will see, this is linked to $\alpha_{\nu_1, \nu_2}$ by detailed balance.

We now impose KMS detailed balance [\autoref{eq:rhoDB}] with respect to a thermal density matrix $\rho_\beta = \frac{1}{Z} e^{-\beta H}$ and  ask how this constrains  $\mathcal{L}$. 

First, we consider the transition part. From a straightforward calculation (\appref{app:DBderivation}), we have
\begin{align}
 (\Gamma_{\rho_\beta}^{-1}\circ \mathcal{T } \circ \Gamma_{\rho_\beta})[\bigcdot] &= \sum_{a\in \mathcal{A}}\sum_{\nu_1, \nu_2\in\bohr} \alpha_{\nu_1, \nu_2}e^{\frac{\beta(\nu_1+\nu_2)}2} \nonumber\\ & \quad\quad\quad\quad \quad \times  A^a_{\nu_1}(\bigcdot)(A^a_{\nu_2})^\dagger,\label{eq:Tconjugatedmain}
\end{align}
Meanwhile, under the assumption that the set of jump operators $\mathcal{A}$ is closed under Hermitian conjugation (i.e. $A_a\in \mathcal{A}$ iff $A_a^\dagger\in\mathcal{A}$), a slightly more involved computation (detailed in \appref{app:DBderivation}) yields
\begin{align}
\mathcal{T}^\dagger[\bigcdot]  =\sum_{a\in \mathcal{A}}\sum_{\nu_1, \nu_2\in\bohr} \alpha_{-\nu_2, -\nu_1} A^a_{\nu_1}(\bigcdot)(A^a_{\nu_2})^\dagger.\label{eq:Tdaggermain}
\end{align}
Comparing \autoref{eq:Tconjugatedmain} and \autoref{eq:Tdaggermain}, we see that  $\mathcal{T}$ is KMS detailed balanced, i.e. $\mathcal{T}^\dagger =  \Gamma_\rho^{-1}\circ \mathcal{T} \circ \Gamma_\rho$, if 
\begin{equation}
\alpha_{-\nu_2, -\nu_1} = \alpha_{\nu_1, \nu_2}e^{\frac{\beta(\nu_1+\nu_2)}{2}}.\label{eq:alphaDBmain}
\end{equation}

While $\mathcal{T}$ satisfies detailed balance on its own, for generic choices of $\alpha_{\nu_1, \nu_2}$ satisfying \autoref{eq:alphaDBmain} the decay and coherent parts  $\mathcal{R}$ and $\mathcal{G}$ do not satisfy detailed balance individually,  but instead are intertwined by the action of $\Gamma_\rho$. However, as first noted by Ref.~\onlinecite{CKG23} (and here summarized in \appref{app:DBderivation}), $\mathcal{R}$ and $\mathcal{G}$ {\it together} satisfy KMS detailed balance, i.e. $(\mathcal{G}+\mathcal{R})^\dagger =\Gamma_\rho^{-1}\circ(\mathcal{G}+\mathcal{R}) \circ \Gamma_\rho$,  if we choose
\begin{equation}
g_{\nu_1, \nu_2}  = -\frac{1}{2i}\tanh\frac{\beta(\nu_1-\nu_2)}{4}\alpha_{\nu_1, \nu_2}\label{eq:gexactmain}. 
\end{equation}

Any set of coefficients $\alpha_{\nu_1, \nu_2}$ and $g_{\nu_1, \nu_2}$ that satisfy the conditions in \autoref{eq:alphaDBmain} and \autoref{eq:gexactmain} generates a Lindbladian via \autoref{eq:TRGdefs} that exactly satisfies KMS detailed balance, and hence has the thermal state $\rho_\beta$ as a fixed point. Two choices are especially salient. The first is to take 
\begin{equation}
    \alpha_{\nu_1, \nu_2} = \delta_{\nu_1, \nu_2}\gamma(\nu_1)\quad {\rm with }\quad \gamma(-\nu) = \gamma(\nu)e^{\beta \nu},
\end{equation} 
in which case  $G=0$. This corresponds to the well-known Davies Lindbladian. The drawback, as noted previously, is that generating such Lindblad dynamics by  evolving local jump operators under a local $H$ requires a time exponentially long in system size (since in order to resolve the smallest Bohr frequencies, we need to evolve until the Heisenberg time).

A different choice, and our main focus, is~\cite{ding2024efficient}
\begin{align}\label{eq:alphaexact}
    \alpha_{\nu_1, \nu_2} &= \hat{f}(- \nu_1) [\hat{f}(- \nu_2)]^*
\end{align}
with
\begin{subequations}\label{eq:exactDBL}
\begin{align}\label{eq:exact_filter}
   \hat{f}(\nu) &= e^{+\beta \nu/4}q(\nu),  \quad {\rm and } \quad  q(-\nu) = q^*(\nu).
\end{align}
Using this form of filter function, the transition and decay parts in \autoref{eq:TRGdefs} can be brought into the form 
\begin{equation}\label{eq:Gibbsexact}
    \mathcal{L}_\beta[\bigcdot] = -i[G,\bigcdot ] +\sum_{a\in \mathcal{A}} L_a^{\pdag}(\bigcdot)L_a^\dagger -\frac{1}{2}\left\{L_a^\dagger L^\pdag_a, \bigcdot \right\},
\end{equation}
where 
\begin{equation}\label{eq:exactDBLcoherent}
   G = \frac{i}{2}\sum_{a\in\mathcal{A}}\sum_{\nu\in\bohr}\tanh\left(\frac{\beta\nu}{4}\right) (L^\dagger_a L_a^\pdag)_\nu 
\end{equation}
and
\begin{equation}\label{eq:Lfdef}
L_a = \sum_{\nu\in \bohr} \hat{f}(-\nu) (A_a)_\nu = \int_{-\infty}^\infty f(t) A_a(t) dt,
\end{equation}
\end{subequations}
with $f(t) = \frac{1}{2\pi}\int_{-\infty}^\infty \hat{f}(\nu)e^{i\nu t} d\nu$ [as can be seen using \autoref{eq:Bohrdef}]. (We have added a subscript $\beta$ to emphasize that \autoref{eq:Gibbsexact} is detailed balanced  with respect to $\rho_\beta$.)  For a suitable choice of $q(\nu)$ (e.g., a Gaussian), the time-domain filter function decays quickly, so we can view  \autoref{eq:Lfdef} as a ``smoothing'' of the time-evolved jump operator. As such, for local $H$, the $L_a$ obtained in this way will be quasi-local due to Lieb-Robinson bounds.

The KMS condition [\autoref{eq:rhoDB}] with respect to $\rho_\beta$ is then conveniently stated as 
\begin{equation}
\rho_\beta^{-1/2} L_a^\pdag \rho_\beta^{1/2} =  L_a^\dagger.
\label{eq:KMSonL}
\end{equation} 
 Ref.~\onlinecite{ding2024efficient} shows that Lindbladians that satisfy KMS detailed balance (i.e., exact Gibbs samplers) can always be written in the form of  \autoref{eq:exactDBL}.

\section{The Local Driving Sampler \label{sec:LDS}}

The form of the exact Gibbs sampler, where quasi-local jump operators are generated by time evolution of strictly local operators, suggests that one might be able to generate such a Lindbladian by leveraging the operator spreading that naturally occurs under Hamiltonian dynamics.
Combining this observation with known results on Lindbladian simulation via Hamiltonian simulation, we show below that we can implement, efficiently and with controlled error, a Lindbladian that reproduces an exact Gibbs sampler {\it up to the coherent part}. 
While the latter is important to {\it exactly} satisfy KMS detailed balance, this is inessential if we require only that the thermal state is an \emph{approximate} steady state \cite{correa2023lamb_shift, CKBG23}. 
We show below in \autoref{sec:proofs} that the Lindbladian that we implement is indeed such an approximate Gibbs sampler with controlled fixed-point error, when assuming that the mixing time does not scale adversely with either system size or protocol parameters. 
To derive such guarantees, the closeness to a Lindbladian which satisfies the KMS-detailed balance condition is essential.
(To sequester the technical portion of the work to a single section, here we only discuss the high-level picture and relegate detailed derivations to where we prove error bounds in \autoref{sec:proofs} and Appendices~\ref{app:bound_magnus} and \ref{app:boundfpe}.)

Our setup, sketched in \autoref{fig:setup}, consists of a system with Hamiltonian $H$ acting on $\nS$ system qubits, coupled to a bath of $\nB$ ancilla qubits, one per jump operator $A_a$, $a=1, 2\,\ldots |\mathcal{A}| = \nB$. 

The Hamiltonian for the coupled system is:
\begin{align}\label{eq:HSB_schroedinger}
H_{\rm SB} &= I_B \otimes H + H_B\otimes I_s \nonumber\\
    &\phantom{=} + \sum_a J\{f(t) B_a^\dagger A_a + f^*(t) B_a A_a^\dagger \}
\end{align}
where $f(t)$ is taken to coincide with the  filter function appearing in the exact sampler [cf. \autoref{eq:exact_filter} and below \autoref{eq:Lfdef}] and $B= \tfrac{1}{2}(X_B  - i Y_B)$, $B^\dagger = \tfrac{1}{2}(X_B + i Y_B)$ are lowering and raising operators for the ancilla, which satisfy $B^2 =  (B^\dagger)^2 =0$ and $B^\dagger\ket{0} = \ket{1}, B\ket{0} =0$.  We will often be interested in the case where the bath Hamiltonian is trivial, but for now allow there to be some nontrivial bath dynamics governed by $H_B$; note however that we will {\it only}  consider the case where the bath is noninteracting, i.e. the bath Hamiltonian $H_B$ does not involve coupling between the ancillas. (The case where the bath interactions are purely diagonal can be shown to be equivalent to this, but may prove useful e.g. in engineering appropriate $f(t)$'s.) 

We now move to the interaction picture with respect to $H + H_B$: the corresponding interaction-picture Hamiltonian is then denoted by a tilde, and consists purely of a system-bath interaction
\begin{align}\label{eq:Hint}
\tilde{H}_{\rm SB}(t) &= J \sum_a \big\{ f(t) B_a^\dagger(t) \otimes A_a(t) \nonumber\\
    &\phantom{=J\sum_a\{} + f^*(t) B_a(t)\otimes A_a^\dagger(t)\big\}
\end{align}
with $\mathcal{O}_B(t) \equiv e^{i H_B t} \mathcal{O}_B e^{-i H_B t}$ for $\mathcal{O} = X, Y$, and $A(t) \equiv e^{i H t} A e^{-i H t}$ as above.

We assume that the $f(t)$ remains operational over some finite time interval $[-T/2, T/2]$, chosen to be symmetric for convenience. Returning to the Schr\"odinger picture, we see that this implements the unitary $e^{-iHT} \tilde{V}$. We complete the cycle by `resetting' the bath to $\rho^0_B = \ket{0}\bra{0}_B$, while implementing a ``rewinding step'' only on the system: a \emph{backwards} time evolution with $H$ for a period $T$. In other words, a single cycle of the time evolution implements the following channel on the \emph{Schrödinger picture} density matrix  $\rho$:
\begin{equation}\label{eq:channelev}
\mathcal{K}[\rho] = \tr_B \left[ \tilde{V} \left(\rho^0_B\otimes {\rho}\right)\tilde{V}^\dagger  \right].
\end{equation}
We emphasize that here, 
\begin{equation}\label{eq:Vexact}
\tilde{V} \equiv \mathcal{T}_t \exp \left( - i\int_{-{T}/{2}}^{T/{2}}d t\, \tilde{H}_{\rm SB}(t) \right)
\end{equation}
is the \emph{interaction-picture} time evolution operator for the composite system comprising the target system and the bath of ancillas, even though  \autoref{eq:channelev} describes the evolution of the {\it Schr\"odinger} density matrix. This explains the point of the ``rewinding'' step: to allow a repeated application of the channel corresponding to evolution under $\tilde{V}$ followed by a bath reset, which is {\it not}  equivalent to simply evolving under $H + V(t)$ for time $T$ followed by a reset.

The idea is that such a channel implements a Liouvillian time evolution over an controllably small interval $\Delta\tau = J^2$ up to an error $O(J^4)$. 
To see this, expand $\tilde{V}$ in powers of $J$ up to third order via the Magnus expansion: we may write 
\begin{align}\label{eq:Vmagnus}
    &\norm{\tilde{V}-\exp(-i \sum_{n=1}^3\tilde{H}_M^{(n)})}_\infty \nonumber\\
    &\quad\quad\quad\quad\quad
    =O\left(
    \nB \left(\frac{JT}{\sigma}\right)^4\exp(\frac{\beta^2}{2 \sigma^2})
    \right), 
\end{align}
as is derived in \autoref{sec:proofs_magnus}. Here, $\tilde{H}_M^{(n)}$ represents the $n^{\rm th}$-order term in the Magnus expansion in the interaction picture, and $\norm{A}_\infty=\sup_{\norm{v}=1} \norm{Av}$ where $\norm{v}=\sqrt{\sum_{i} |v_i|^2}$ denotes the spectral norm.

We now insert the final expression \autoref{eq:Vmagnus} for $\tilde{V}$ into \autoref{eq:channelev} and expand the exponential. Owing to the assumption that the ancilla qubit is reset after each cycle, we can restrict our attention to terms in the expansion that involve $H_F^{(m)} \cdot H_F^{(n)}$ with $m+n$ even, which enter at $O(J^{m+n})$. Accordingly, we have
\begin{align}\label{eq:Kexpanded}
    \mathcal{K}[\rho] &= \rho + \tr_B \left[ \tilde{H}_M^{(1)} \left(\rho^0_B \otimes \rho\right)\tilde{H}_M^{(1)}  \right] \nonumber\\
    &\phantom{=} -\frac{1}{2} \tr_B \left[ \left\{\left(\tilde{H}_M^{(1)}\right)^2, \left(\rho^0_B \otimes \rho\right)\right\}  \right] \nonumber\\
    &\phantom{=} -i \tr_B [\tilde{H}_M^{(2)}, \rho^0_B \otimes \rho] + O(\nB J^4). 
\end{align}

A straightforward but tedious calculation, presented in \appref{app:bound_magnus}, now yields the fact that
\begin{equation}
    \mathcal K[\rho] = e^{J^2 \mathcal L_T}[\rho] + O(\nB J^4)
\end{equation}
where the effective Lindbladian $\mathcal L_T$ can be written in a very appealing form: we have
\begin{subequations}\label{eq:LDScombined}
    \begin{equation}\label{eq:LDS}
\mathcal{L}_T[\bigcdot] =  -i [H_{{\rm LS}}, \bigcdot] + \mathcal{L}_{{\rm diss};T}[\bigcdot],   
\end{equation}
where we have separated the $f$-smoothed Lindblad dynamics into a purely dissipative evolution
\begin{equation}
\mathcal{L}_{{\rm diss};T}[\bigcdot] = \sum_a L_{a;T}\bigcdot L_{a;T}^\dagger - \frac{1}{2}\{ L_{a;T}^\dagger L_{a;T}, \bigcdot\}
\end{equation}
with the jump operator. 
\begin{equation}\label{eq:finiteTsmooth}
    L_{a;T} \equiv \int_{-T/2}^{T/2} f(t) A_a(t) dt,
\end{equation}
and a coherent Lamb shift contribution governed by the Hamiltonian
\begin{align}
H_{{\rm LS}; T} &= -\frac{1}{2i} \int_{-{T}/{2}}^{T/{2}}d t_1 \int_{-{T}/{2}}^{T/2}d t_2\, f^*(t_2) f(t_1) \nonumber\\
&\phantom{--}\times\text{sgn}(t_1-t_2) \sum_a A_a^\dagger(t_2) A_a(t_1)\label{eq:HLStimedomain}\\
& = \sum_{a\in \mathcal{A}}\sum_{\nu_1, \nu_2}   h^{(T)}_{\nu_1, \nu_2}(A^a_{\nu_2})^\dagger A^a_{\nu_1} \label{eq:BohrLS}
\end{align}
where 
\begin{align}
    h^{(T)}_{\nu_1, \nu_2} &= -\frac{1}{2i} \int_{-{T}/{2}}^{T/{2}}d t_1 \int_{-{T}/{2}}^{T/2}d t_2\, f(t_1) f^*(t_2)
     \nonumber\\
&\phantom{--}\times\text{sgn}(t_1-t_2) e^{i\nu_1 t_1 - i \nu_2 t_2}\label{eq:LSkernel}
\end{align}
\end{subequations}
Note that the transition, decay, and coherent (Lamb shift) contributions each respectively arise from the three different traces over the bath in \autoref{eq:Kexpanded}. 

Observe that $\mathcal L_T$ has almost precisely the form of the exact Gibbs sampler in  \autoref{eq:exactDBL}, up to a finite-time truncation of the $f$-filtered Heisenberg evolution, but with a distinct coherent term: instead of the form $G$ required by the exact sampler [\autoref{eq:exactDBLcoherent}], the coherent evolution generated by the local drive is a ``physical'' Lamb shift \cite{correa2023lamb_shift}.

This second distinction between the exact and local driving samplers is most easily seen by taking $T\to \infty$ in the latter; the resulting Lindbladian has transition and decay parts that take the the form of \autoref{eq:Tdef} and \autoref{eq:Rdef} with $\alpha_{\nu_1, \nu_2}$ given  by \autoref{eq:alphaexact},
but with a different coherent part; instead of a kernel $g_{\nu_1, \nu_2}$ in \autoref{eq:Gdef} that satisfies the condition \autoref{eq:gexactmain} for exact detailed balance, we instead have
\begin{equation}
g_{\nu_1,\nu_2}\to h_{\nu_1,\nu_2} \equiv 
\lim_{T\to\infty} h^{(T)}_{\nu_1,\nu_2}.
\end{equation}

We have so far made no specific choice of filter function $f(t)$. Henceforth, we make the convenient and specific choice of a shifted Gaussian \cite{ding2024efficient},
\begin{equation}\label{eq:chosenfilterfunction}
    f(t) = \sqrt{\frac{2}{\pi\sigma^2}}e^{-\frac{2}{\sigma^2}\left(t-\frac{i\beta}{4}\right)^2},
\end{equation}
corresponding to taking $q(\nu) = e^{-\frac{(\sigma \nu)^2}{8}}$ in \autoref{eq:exact_filter}. This fixes both the exact KMS-detailed balanced and the local-driving Lindbladians that we consider in the remainder of this paper. 
In the next section, we bound the error in approximating the former by the latter.

\section{Bounds on the Accuracy\label{sec:proofs}}

\begin{figure*}
    \centering
    \includegraphics[width=\textwidth]{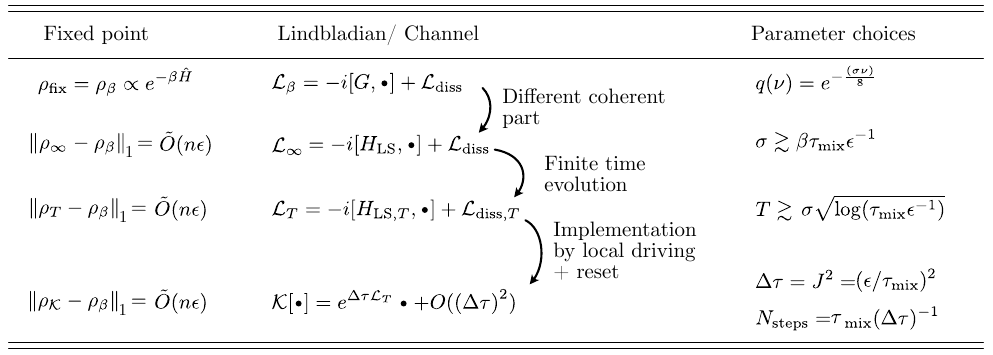}
    \caption{Summary of different approximations made during the derivation of the local driving sampler protocol.
    Overall, the setup in \autoref{fig:setup}, with the parameter choices indicated in the rightmost column above, implements a channel $\mathcal K$ that approximates an exact Gibbs sampler and hence has an approximately thermal fixed point. As discussed in \autoref{sec:optimal_accuracy}, the parameter choices indicated in the last column may be non-trivial.}
    \label{fig:bound_summary}
\end{figure*}

In this section, we summarize the different error sources and their contributions to the final state when using the setup described in \autoref{sec:LDS} to prepare an approximate thermal state of a Hamiltonian $H$. We use these results to derive an upper bound on the Hamiltonian simulation time necessary to prepare the thermal state $\rho_{\beta}$ up to some error per qubit $\epsilon$ in trace norm. This will also clarify  the conditions under which this error can be made arbitrary small, and how to estimate the minimal error that can be achieved by our protocol.

We summarize the different approximations made in \autoref{fig:bound_summary}.
The total error of implementing the local driving protocol as opposed to the exact sampler in \autoref{eq:exactDBL} stems, roughly speaking, from two sources. First, the physical protocol implements an finite time evolution with a Lindbladian $\mathcal L_T$, up to an error which is controlled by the truncation error of the third-order Magnus expansion in the interaction picture \cite{sharma2024simulation}. As explained below, this in turn is governed by the time step $\Delta\tau$ in the discretization of the target Lindblad evolution, which in effect scales as the second power of the system-bath coupling $J$. 
The second source of error is the fact that even if we could perfectly realize the Lindbladian $\mathcal L_T$, it has a steady state $\rho_{\rm SS}$ (i.e. $\mathcal L_T[\rho_{\rm SS}] = 0$) which is only approximately equal to the thermal state, i.e. $\rho_{\rm SS} \approx \rho_{\beta}$. This \emph{fixed-point error} is ultimately controlled by the parameter $T$ of $\mathcal L_{T}$, the width $\sigma$ of the filter function, and the mixing time $\tmix$. 
 These quantities enter the error bound in two ways. Clearly, $T$ affects the extent to which the finite-$T$ smoothing of jump operators in \autoref{eq:finiteTsmooth}  approximates that of the exact sampler in \autoref{eq:Lfdef}. Less obviously, $\sigma$, $T$, $\tmix$ together govern the modification of the fixed point due to the ``wrong'' coherent evolution, via their influence on the ``secular approximation'' discussed in \autoref{sec:fixed_point_err}. Note that the quantities entering the fixed-point error are not independent, as the mixing time in general will depend on both $\sigma$ and $T$. As discussed in \autoref{sec:resources}, if this dependence is too strong, it can limit the minimal accuracy achievable by the protocol.

We bound the fixed-point error in terms of the trace norm:
\begin{align}
    \norm{\rho-\sigma}_1=\tr\sqrt{(\rho-\sigma)(\rho-\sigma)^\dagger}.
\end{align}
The trace norm gives an upper bound on the error in observables, as $\tr A(\rho-\sigma)\leq \norm{A}_\infty\norm{\rho-\sigma}_1$.
Moreover, in order to prove the bounds, we make use of the Schatten p-norms defined as
\begin{align}
    \norm{\rho}_p=(\tr (\sqrt{ \rho \rho^\dagger})^p)^{1/p}
\end{align}
and the induced super-operator norms
\begin{align}
    \norm{K[\bigcdot]}_{p\rightarrow p}=\sup_{\norm{X}_p=1} \norm{K[X]}_p.
\end{align}

\subsection{Lindbladian Simulation Error\label{sec:proofs_magnus}}
As we advertised, our protocol implements a Lindbladian $\mathcal{L}_T$ approximately by performing a time-dependent Hamiltonian simulation followed by a reset; the key is that for a suitable choice of $f(t)$, the drive approximates the jump operators that enter the exact sampler. 
This approximation stems from replacing the time-ordered integral $\Tilde V$ in \autoref{eq:Vexact} by its third-order Magnus expansion (cf. \autoref{sec:LDS}). Therefore, to bound this error we can leverage recent results on Hamiltonian simulation in the interaction picture \cite{sharma2024simulation} which show that
\begin{equation}
    \norm{\Tilde V - \exp[-i\left( \sum_{k=1}^3\Tilde H_{\rm SB}^{(k)} \right)]}_1 = O\left( \nB \left(d J T\right)^4 \right)\label{eq:Vmagnusmain}
\end{equation}
where $\Tilde H_{\rm SB}^{(k)}$ is the $k$th-order term in the Magnus expansion of $\tilde H_{\rm SB}$, and $d = \max_t \abs{f(t)}$. 
We use this, in \appref{app:bound_magnus}, to show that for all $\rho$,
\begin{multline}\label{eq:bound_magnus_channel}
    \norm{e^{J^2 \mathcal L_{T}}[\rho] - \mathcal K[\rho]}_{\infty} \\= O\left(J^4 \left(\nB \left(\frac{T}{\sigma}\right)^4+\nB^2\right)\e^{\frac{\beta^2}{2 \sigma^2}}\right) 
\end{multline}

where $\mathcal L_T$ is given exactly by \autoref{eq:LDScombined}.

In summary, our protocol implements a Lindbladian time step of size $\Delta\tau = J^2$ up to an error $O((\Delta\tau)^2)$, which allows one in principle to implement evolution with controlled error for arbitrary long times.

\subsection{Fixed-point Error\label{sec:fixed_point_err}}

Accepting the fact that our protocol implements $\mathcal L_T$ up to controllable error, it remains to bound the `fixed-point error', i.e. the difference of the steady state of the dynamics generated by $\mathcal L_T$ and the thermal state $\rho_\beta$. The fixed-point error again has two distinct sources, that is (i) the fact that we implement the dissipative part of the exact Gibbs sampler only up to some finite evolution time $T$, and (ii) the fact that $\mathcal L_{T}$ even for $T\to\infty$ implements the ``wrong'' coherent part of the evolution. In turns out that both of these lead to bounded errors per time step. The total error is then the error per time step multiplied by the mixing time, which governs the total Lindbladian evolution time necessary to reach the thermal state.

For a choice of filter function $f(t)$ that decays quickly on some finite time scale $\sigma$, restricting $\mathcal{L}_T$ to finite $T$ incurs only a minor error in the Lindbladian evolution in the sense that $\mathcal L_{\infty}$ and $\mathcal L_{T}$  are close in superoperator norm for $T \gg \sigma$. This then bounds the difference between their respective fixed points,
\begin{align}\label{eq:bound_finite_time}
    \norm{\rho_T - \rho_{\infty}}_1 &\leq 4\,\tmix(\mathcal{L}_T)\, \norm{\mathcal L_{T} - \mathcal L_{\infty}}_{1-1} \nonumber\\
    &\leq \frac{24 \sqrt{2}}{\sqrt{\pi}} 
       \left(\frac{\sigma}{T}\right) \e^{\frac{\beta^2}{4\sigma^2}} \e^{-\frac{T^2}{2\sigma^2}} t_{\text{mix}}(\mathcal{L}_T)
\end{align}
where the first inequality is Lemma II.1 of Ref. \onlinecite{CKBG23}, and the second, derived in \appref{app:bound_finite_time}, follows for the specific choice of a Gaussian filter function with standard deviation $\sigma$. Here and below, we define the mixing time  of a Lindbladian as the smallest time $\tmix$ for which, given any $\rho_1, \rho_2$, we have
\begin{equation}\label{eq:mixing_time}
 \norm{e^{\tmix\mathcal{L}}[\rho_1 - \rho_2]}_1 \leq \frac{1}{2} \norm{\rho_1-\rho_2}_1.
\end{equation}

Given the error bound in \autoref{eq:bound_finite_time}, it remains to bound the difference between the fixed point $\rho_{\infty}$ of $\mathcal L_{\infty}$ and the thermal state. While this is technically the most involved step of the whole derivation it follows from a clear intuition. Compared to the exact sampler in \autoref{eq:exactDBL}, the non-truncated local driving sampler $\mathcal L_{\infty}$ has an `added' coherent part $B = H_{\rm LS} - G$. Now, writing $B$ in terms of energy transfers corresponding to the Bohr frequencies of $H$, 
\begin{equation}
    B = \sum_{\nu_1, \nu_1 \in \bohr} b_{\nu_1, \nu_2} (A_{\nu_2})^\dagger A_{\nu_1}
\end{equation}
we can show that if the matrix $b_{\nu_1, \nu_2}$ is fast decaying for $\abs{\nu_1 - \nu_2} \gg \mu$ and some $\mu > 0$, then the operator $B$ almost commutes with $\rho_{\beta}$. Therefore, adding it to $\mathcal L_{\beta}$ as a coherent evolution should leave the fixed point almost unchanged. The challenge is to quantify the ``almost'', which technically  requires employing the so-called ``secular approximation'' \cite{breuer_petruccione2002,Mozgunov2020completelypositive,correa2023lamb_shift,CKBG23}. In our case, we show in \appref{app:bound_secular}, that
\begin{equation}\label{eq:bound_secular}
    \norm{\rho_{\infty} - \rho_{\beta}}_1 = \Tilde O\left(
    \nB\,
    \frac{\beta}{\sigma}\,
    \max(\tmix(\mathcal L_{\infty}), \tmix(\mathcal L_{\beta})
    \right)
\end{equation}
where $\Tilde O$ denotes an upper bound up to (poly)logarithmic corrections.
As long as the mixing time $\tmix$ does not grow super-linearly with the width of the filter function in the time domain, $\sigma$, this part of the fixed-point error is hence controlled by $\sigma$ compared to the inverse temperature $\beta$.

Combining the two sources of fixed-point error, we find

\begin{align}\label{eq:bound_fixed_point}
    \norm{\rho_T - \rho_{\beta}}_1 &= \Tilde O\left(
        \nB 
        \left(\left(\frac{\sigma}{T}\right) \e^{\frac{\beta^2}{4\sigma^2}} \e^{-\frac{T^2}{2\sigma^2}} 
            + \frac{\beta}{\sigma} 
        \right)
        \tmixmax
        \right)
\end{align}

where $\tmixmax \equiv \max(\tmix(\mathcal L_T), \tmix(\mathcal L_{\infty}), \tmix(\mathcal L_{\beta}))$. This
informs the choice of filter function and finite-time truncation: we want to choose $T \gg \sigma \gg \beta$ [cf. \autoref{eq:LDS_params}].

\subsection{Resource Estimate}\label{sec:resources}

\begin{figure*}[t!]
    \centering   \includegraphics{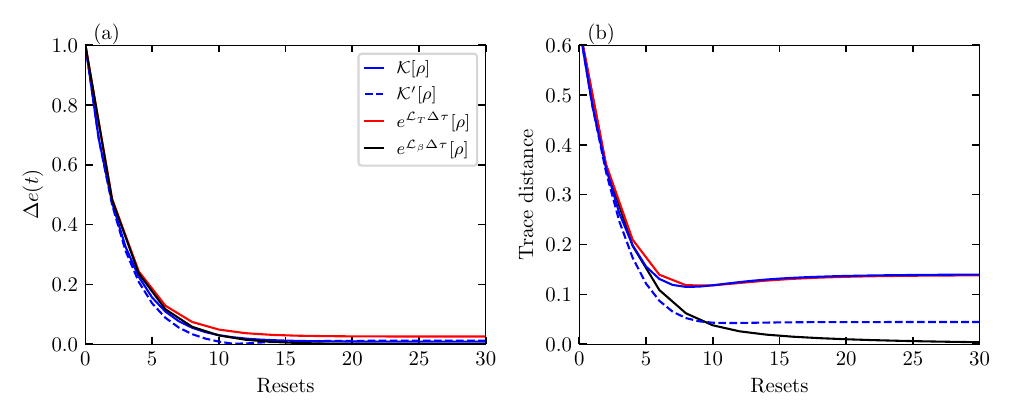}
    \caption{Comparison of the time evolution of thermal state preparation for the mixed-field Ising model~\autoref{eq:MFI} under the exact Gibbs sampler $\cL$ [cf.~\autoref{eq:Gibbsexact}] (black), the Lindbladian $\cL_T$ [cf.~\autoref{eq:LDS}] (red), the local driving protocol $K[\rho]$~[cf.~\autoref{eq:channelev}] (blue), and the protocol $K^\prime[\rho]$ without rewinding~(dashed blue) starting from the maximally mixed state $\rho_0 = 2^{-\nS}\mathds{1}$.  
    Parameters: $\beta = 1.0$, $\sigma = 0.5$, $J=0.5$ , $n_S = 8$, $T=6\sigma$. (a) Time evolution of the error in the average energy density $\Delta e(t)=\frac{\langle H-H_\beta\rangle}{\langle H_\beta \rangle}$. The energy density approaches the exact result within less than two percent.  
    (b) Evolution of the trace distance to the Gibbs state $\rho_\beta$. The trace distance decays, indicating convergence toward a state close to the Gibbs state. 
}
    \label{fig:Numerics}
\end{figure*}

Using the error bounds derived above, we can estimate the total cost of preparing a thermal state $\rho_{\beta}$ up to some precision $\epsilon$. For that, we want both the fixed-point error \autoref{eq:bound_fixed_point}, as well as the total accumulated simulation error \autoref{eq:bound_magnus_channel} to be $\Tilde O(\nB \epsilon)$.

Assuming for a moment that all quantities appearing in the error bounds are independent, then for a given target accuracy $\epsilon$ (per qubit), we choose
\begin{subequations}\label{eq:LDS_params}
\begin{align}
    \sigma &= \frac{\beta \tmixmax}{\epsilon} \\
    T &= \sigma \sqrt{\log \frac{\tmixmax} {\epsilon}} 
    = \frac{\beta \tmixmax}{\epsilon}\sqrt{\log \frac{\tmixmax}{\epsilon}}
\end{align}
\end{subequations}
which, upon substitution into \autoref{eq:bound_fixed_point}, yields
\begin{align}
    \norm{\rho_T - \rho_{\beta}}_1
    &= \Tilde O\left(\nB \epsilon \,(1+ \sqrt{\log\epsilon^{-1}}) \right) \\
    &= \Tilde O(\nB\epsilon).
\end{align}

Further, the total accumulated error during a Lindbladian time evolution of time $\propto\tmix\equiv\tmix(\mathcal L_{T})$ is $O(\nB\tmix J^2) = O(\nB\epsilon)$ if $J^2 = \tilde{O}(\epsilon \tmix^{-1})$ and hence we need $N_{\rm step} = O(\tmix^2 \epsilon^{-1})$ elementary steps of the protocol to reach the steady state.
The total Hamiltonian evolution time necessary to prepare the thermal state $\rho_\beta$ up to error $\Tilde O(\nB\epsilon)$ in trace norm then scales as
\begin{subequations}\label{eq:resource_estimate}
\begin{align}
    T_{\rm tot} &= O(N_{\rm step} T) \\
        &= \Tilde O\left(\frac{\beta \, \tmix^2 \, \tmixmax}{\epsilon^2}\sqrt{\log \frac{\tmixmax}{\epsilon}}\right) \\
        & = \Tilde O\left(\frac{\beta \, \tmix^2 \,\tmixmax}{\epsilon^2}\right).
\end{align}
\end{subequations}
This remains polynomial in system size 
even if we demand that the fixed-point is reached up to an $O(1)$ total error [in this case the error per bath qubit must be $\epsilon^{-1} \sim \tilde{O}(\nB)\sim \tilde{O}(\nS)$], assuming a polynomial-in-system size mixing time.
In practice, the method is most effective if the mixing time has a mild (e.g. logarithmic) scaling with $\nS$, and $\epsilon$ can be chosen to be some $O(1)$ number, which we expect to be sufficient in many cases to obtain e.g. local observables in the thermal state up to an $\Tilde O(\epsilon)$ error.

The mixing time is in general controlled by spectral properties of the Lindbladian \cite{kastoryano2012quantum_sobolev, temme2010divergence}. Although there exist Hamiltonians for which any dynamics with (quasi-)local jump operators can be shown to have a mixing time that is exponentially slow in system size \cite{rakovszky2024bottleneck, placke2024slow_mixing,garmarnik2024slow_mixing,anschuetz2024glassiness},
fast mixing has been established for many models and temperature regimes of interest \cite{znidaric2015relaxation,bardet2023rapid_mixing,kochanowski2024rapid_mixing,tong2024poly_mixing,ding2024poly_mixing,smid2025poly_mixing,rouze2024efficient,rouze2024optimal}.

The discussion above is true even in the case where the mixing time $\tmixmax$ depends on $\sigma$ and $T$, as long as we can make the choice of parameters in \autoref{eq:LDS_params}. However, if $\tmixmax$ depends on $\sigma$, this choice becomes nontrivial, or even impossible, for some $\epsilon$. Because of this, in general there exists an minimal relative error $\epsilon^*$ that can be achieved, which we discuss in the following.

\subsection{Optimal Accuracy\label{sec:optimal_accuracy}}

While the derived bounds on the fixed-point error apply to arbitrary choices of Hamiltonians, temperatures, and parameter $\sigma$ in the filter function, it remains an open question to understand the extent to which the fixed point $\rho_\beta$ can be approached by the channel $K[\rho]$: in other words, to establish the $\epsilon$ for which the parameter choice in \autoref{eq:LDS_params} is possible.
The choice is nontrivial because the mixing time $\tmixmax$ in general depends on the Lindbladian, and hence the parameters $\sigma$ and $T$.

The dependence on $\sigma$ in particular is problematic since it can be very strong. 
This is because increasing the temporal filter width $\sigma$ correspondingly decreases the width of the filter in frequency space, 
and transitions with energy $\Delta E$ are exponentially suppressed if $\Delta  E \gg \sigma^{-1}$.
In systems with a discrete spectrum (e.g. few-body or non-interacting), and without fine-tuning the temperature, one thus expects the mixing time to diverge \emph{exponentially}, as soon as the inverse filter width $\sigma^{-1}$ is smaller than the level spacing. The same problem also arises in the arguably more interesting setting of gapped systems at temperatures smaller than the gap (recall that we chose $\sigma \gg \beta$).
In these cases, the fixed-point error cannot be arbitrarily reduced using the channel $K[\rho]$, and there exists an optimal choice $\sigma^*$ that minimizes it.

The dependence of the mixing time on $\sigma$ becomes more involved in interacting systems at intermediate temperatures. In this case, the level spacing is expected to be exponentially small, and there are possible transitions for any $\sigma$. Still, even in this case numerical results for the mixed field Ising model~(\appref{subsec:Mixingtimeanalysis}) show an exponential increase of the mixing time with $\sigma$. This suggests that even in this case, there exists a finite $\sigma^*$ which minimizes the fixed-point error.

We note that the strong dependence of the  mixing time $\tmixmax$ on the parameter $\sigma$ limits not only our approximate protocol, but also poses problems 
for exact samplers, where $\sigma=\beta$ is a common choice of parameters \cite{CKG23,ding2024efficient}. Specifically, while such a choice does not limit the {\it accuracy} of exact samplers (obvious, given the name), it does limit their {\it efficiency} in preparing the thermal state from  arbitrarily  initial conditions. 
Developing a better understanding of the dependence of the mixing time on $\sigma$ is therefore an important task for future work, both to find regimes of best applicability and also how to avoid this dependence altogether. 
We note that since an  early version of this work appeared in preprint form, there have been two notable proposals in this direction \cite{lloyd2025quantumthermalstatepreparation,ding2025endtoendefficientquantumthermal}.

Despite these obvious drawbacks, we note that our protocol numerically yields good performance in certain regimes, as we explore for a simple example in \autoref{sec:numerics}. 

\subsection{Reduction to Quasiparticle Cooling and the ``Rewinding'' Step}

We now comment on the precise link between our work, and the quasiparticle cooling protocol of Refs.~\onlinecite{mi2024quasiparticle,lloyd2024quasiparticle}. Relative to the latter, our  protocol [see \autoref{fig:setup}] implements an additional backwards time evolution after each step. As noted above, this step is what allows us to build a channel that involves repeated application of the interaction-picture unitary $\tilde{V}$. However, evidently this step is superfluous if for each cycle $[\rho(t), H]=0$ at a time $t$ after the application of $\tilde{V}$. This in essence is the `quasiparticle assumption' at the heart of Refs.~\onlinecite{mi2024quasiparticle,lloyd2024quasiparticle}.

However, we note that, since it involves forward time evolution under $-H$, the rewinding step is potentially tricky to implement in practice for analog simulators, and may be costly in any case as it doubles the duration of the protocol. However, we show (in \appref{app:rewindingerror}) that even in the absence of any quasiparticle assumption, dropping the rewinding step only increases the fixed-point error by an additional factor of $\tmix$, viz.
\begin{equation}\label{eq:rewindingerror}
    \norm{\rho_{T}' - \rho_{\beta}} \leq 4\delta \tmix
\end{equation}
where $\delta$ is the error of the protocol \emph{with} rewinding in \autoref{eq:bound_fixed_point}. This leads to an additional factor of $\tmixmax$ in the cost estimate in \autoref{eq:resource_estimate}.
While this can be severe if the mixing time scaling is particularly adverse, the total Hamiltonian time evolution cost nevertheless remains polynomial in system size for any $H$ for which the full protocol has ${\rm poly}(n)$ scaling.

\section{Numerical Case Study\label{sec:numerics}}

To illustrate the feasibility of our protocol, we numerically simulate our protocal as discrete-time quantum channel, and compare the resulting evolution to  the dissipative evolution under the Lindbladians $ \cL_T $ [cf.~\autoref{eq:LDS}] (to confirm that we approximately simulate it) and the exact Gibbs sampler $ \cL_\beta$ [\autoref{eq:Gibbsexact}] (whose simulation is our objective). 

As a concrete example, we consider thermal state preparation for the mixed-field Ising model:
\begin{align}\label{eq:MFI}
    H=\sum_i Z_iZ_{i+1}+g X_i+ h Z_i.
\end{align}
We set $g=0.9045 $ and $ h=0.809$ , parameters for which the model is known to thermalize rapidly~\cite{Kim2013Ballistic}.

To simulate the Lindbladian dynamics, we numerically solve the Lindblad equation using the adaptive Runge--Kutta method of order 5(4)~\cite{DORMAND198019}.
The initial density matrix $ \rho $ is taken to be the maximally mixed ensemble.

The results are shown in Fig.~\ref{fig:Numerics}, where we present data for $\beta=1.0$, $\sigma=0.5$, $J=0.5$, $T=6\sigma$ and a system size of $\nS=8$ qubits. The evolution under the exact Gibbs sampler is shown in black, while the results for the Lindbladian $\cL_\infty$ [cf.~\autoref{eq:LDScombined}] are shown in red and under the local driving protocol $\mathcal{K}$ are shown in blue~[cf.~\autoref{eq:channelev}].
We note that we are in a regime $\beta \gtrsim \sigma$ where the error bounds in our protocol are not necessarily optimal.
Results corresponding to evolving with the Lindbladian $\cL$ or $\cL_\infty$ only are shown in dashed lines. In the Schroedinger picture, this corresponds to the case where reverse time evolution $U=e^{i Ht}$ is applied after the time-dependent system-bath interaction.

In Fig.~\ref{fig:Numerics}(a), we display the dynamics of the average energy, divided by the energy density of the thermal state
\begin{align}
    \Delta e(t)=\frac{\langle H-H_\beta\rangle }{H_\beta},
\end{align}
with $H_\beta=\Tr(H\rho_\beta)$ being the average energy of the Gibbs state. 

In both cases, the energy rapidly converges to the thermal value within less than 2 percent.

Due to the deviations from an exact Gibbs sampler, the results converge to a state close to the Gibbs state. This is shown in 
Figure~\ref{fig:Numerics}~(b), where the trace distance with respect to the thermal state $ \rho_\beta $ is displayed. The trace distance quickly decays, demonstrating rapid convergence to the steady state. 

These results indicate that the local driving sampler yields a good approximation of the exact Gibbs sampler and provides evidence for the effectiveness of our protocol for thermal state preparation.

An open interesting question is the effect of the rewinding step of the protocol $K[\rho]$. In order to investigate its effect, we consider the channel $K^\prime[\rho]$, which is obtained by removing the rewinding step
\begin{align}
    K^\prime[\rho]=\tr_B \left[ e^{-iH T}\tilde{V} \left(\rho^0_B\otimes {\rho}\right)\tilde{V}^\dagger  e^{iHt}\right].
\end{align}
The results for this protocol are indicated by a dashed blue line in Fig.~\ref{fig:Numerics}(a). It turns out that the accuracy of these results improves in comparison to the protocol $K[{\rho}]$. Understanding the reason for this improvement remains an open question for future work.

\section{Concluding Remarks\label{sec:conclusions}}

In this work, we have demonstrated that it is in principle possible to prepare approximate thermal states with rigorously controlled error bounds using ingredients commonly available in the current generation of quantum devices: implementation of a local Hamiltonian, time-dependent local driving, and the ability to reset ancillas. The local driving protocol we introduce and study {\it exactly} simulates the dissipative portions of an exact Gibbs sampler (whose steady state is the thermal density matrix $\rho_\beta$), but rather than the specifically tailored coherent evolution of the latter, involves a physical Lamb shift. 

Our bounds explicitly depend on the mixing time of the approximated Lindbladians, which in turn is controlled  by the properties of the filter functions. Understanding this dependence is therefore crucial to estimate the accuracy to which our approximate protocol can approximate a thermal state, and indeed is important more generally in determining the efficiency of exact Gibbs samplers, if not their accuracy. Since mixing times must often be determined on a case-by-case basis, this represents a challenging question for the future --- although early results in this direction suggest that at least a measure of generality may be possible~\cite{lloyd2025quantumthermalstatepreparation,ding2025endtoendefficientquantumthermal}.

A notable feature of our proposal is that it does not require complicated ``quantum programming'' such as the block-encoding of operators. As such, it can be readily adapted to analog quantum simulators~\cite{QSimReview}, such as those involving ultracold atoms, for which the most nontrivial obstacles will likely be a convenient implementation of the ancilla reset step (essential) and the backward time evolution (desirable). 

The ready applicability of our protocol is underscored by the fact that for a suitable choice of jump operators, it resembles ``quasiparticle cooling'' Refs.~\cite{mi2024quasiparticle, lloyd2024quasiparticle}; this also indicates that (modulo mixing time considerations) the existence of an approximate quasiparticle description is not a necessary condition for such state preparation methods.

The key distinction from the protocols in Ref.~\cite{mi2024quasiparticle, lloyd2024quasiparticle} is the requirement of backward time evolution, which is essential in order that the dissipation match that of the exact Gibbs sampler. From a practical standpoint, eliminating this rewinding step is highly desirable, as it doubles the simulation time and may be challenging to implement in experimental setups. 
Surprisingly, our numerics indicate that removing the rewinding step improves the performance of the protocol.
Formally, this loosens the accuracy bound by a power of the mixing time, so future work~\cite{Hahn2025Rewinding} will explore whether this can be done without compromising the  performance of the protocol in general.

Clearly, it would be highly desirable to develop similar readily-implementable protocols that more closely approximate the exact coherent term of Ref.~\cite{CKG23, ding2024efficient}. Possible routes to this might involve a judicious choice of forward and backward time evolution to ``echo out'' spurious Lamb shift terms, and similar techniques of quantum control. A more careful study of the optimal parameter choice for efficient state preparation --- especially one organized around physical properties of the target state --- as well as a more principled understanding of mixing times for local driving samplers remain important directions for future work. Finally, it would be interesting to ask how these methods compare operationally with others for controlled approximation of thermal observables on quantum or even classical~\cite{WhiteMinimally, kuwahara2016,Stoudenmire_2010,Rigol2006Numerical} computers.

\begin{acknowledgments}
We thank C.-F. Chen for useful discussions on the secular approximation and Minh Tran for helpful correspondence on  Ref.~\onlinecite{sharma2024simulation}, as well as Jerome Lloyd, Dmitry Abanin, and Lin Lin for insightful discussions and engaging correspondence. We further thank Pieter Claeys for helpful discussions.
We acknowledge support from a Leverhulme Trust International Professorship grant (Award Number:
LIP-2020-014, for a Leverhulme-Peierls Fellowship at Oxford to DH, BP),  the UKRI under a  Frontier Research Guarantee (for an ERC Consolidator Grant) EP/Z002419/1 (SAP), and the Alexander von Humboldt foundation through a Feodor-Lynen fellowship (BP).

\end{acknowledgments}

\begin{appendix}
\begin{widetext}

\section{Structure of the Exact Gibbs Sampler from KMS Detailed Balance}
\label{app:DBderivation}
In this section, we provide a detailed derivation of the constraints imposed on the detailed balance Lindbladian by detailed balance, following Ref.~\onlinecite{CKG23}. Our starting point is the Bohr-frequency representation \autoref{eq:TRGdefs}. We will assume (as in the main text) that $\mathcal{L}$ consists either of Hermitian jump operators, or else is `paired' such that  $A^{\pdag}_a$ appears in the set of jump operators $\mathcal{A}$  if and only if $A_a^\dagger$ also appears in $\mathcal{A}$. The reason for this will become apparent below.

Before proceeding, recall that we have already introduced  the ``weighting'' operator $\Gamma_\rho$ involving conjugation with powers of any full-rank state $\rho$:
\begin{equation}
\Gamma_{\rho}[\bigcdot] := \rho^{1/2}(\bigcdot)\rho^{1/2},
\end{equation}
It is convenient to also define a second such operator,
\begin{equation}
\Lambda_{\rho}[\bigcdot] :=\rho^{-1/2} (\bigcdot) \rho^{1/2},
\end{equation}
such that for a Hermitian operator $X$ we have $\Gamma_{\rho}[X]^\dagger =\Gamma_{\rho}[X]$ and $\Lambda_\rho[X]^\dagger = \Lambda_\rho^{-1}[X]$.  

First, we consider the detailed balance condition on the transition part $\mathcal{T}$; a short calculation gives us that
\begin{align}
 (\Gamma_{\rho_\beta}^{-1}\circ \mathcal{T } \circ \Gamma_{\rho_\beta})[\bigcdot] &= \sum_{a\in \mathcal{A}}\sum_{\nu_1, \nu_2\in\bohr} \alpha_{\nu_1, \nu_2} {\rho_\beta}^{-1/2} A^a_{\nu_1}{\rho_\beta}^{1/2}(\bigcdot) {\rho_\beta}^{1/2}(A^a_{\nu_2})^\dagger{\rho_\beta}^{-1/2} \nonumber\\
 &= \sum_{a\in \mathcal{A}}\sum_{\nu_1, \nu_2\in\bohr} \alpha_{\nu_1, \nu_2} e^{\frac{\beta(\nu_1+\nu_2)}2}  A^a_{\nu_1}(\bigcdot)(A^a_{\nu_2})^\dagger.\label{eq:Tconjugated}
\end{align}
where we used the fact that $\rho^{-s} A_\nu \rho^{s} = e^{\beta s \nu } A_\nu$ and hence $\rho^{s} (A_\nu)^\dagger \rho^{-s} = e^{\beta s \nu } (A_\nu)^\dagger$, which follow from  the Bohr frequency representation \autoref{eq:Bohrdef}.

On the other hand, by using $\text{tr}\{\mathcal{O} \mathcal{T}[\rho]\} = \text{tr}\{\rho \mathcal{T}^\dagger[\mathcal{O}]\}$ to determine $\mathcal{T}^\dagger$, we have
\begin{align}
\mathcal{T}^\dagger[\bigcdot] &= \sum_{a\in \mathcal{A}}\sum_{\nu_1, \nu_2\in\bohr} \alpha_{\nu_1, \nu_2} (A^a_{\nu_2})^\dagger (\bigcdot) A^a_{\nu_1} \nonumber\\
&= \sum_{a\in \mathcal{A}}\sum_{\nu_1, \nu_2\in\bohr} \alpha_{\nu_2, \nu_1} (A^a_{\nu_1})^\dagger (\bigcdot) A^a_{\nu_2} &\text{(Relabeling $\nu_1\leftrightarrow \nu_2$)} \nonumber\\
&=\sum_{a\in \mathcal{A}}\sum_{\nu_1, \nu_2\in\bohr} \alpha_{-\nu_2, -\nu_1} (A^a_{-\nu_1})^\dagger (\bigcdot) A^a_{-\nu_2} & \text{(Since if $\nu\in \bohr$ then $-\nu \in \bohr$)}\nonumber\\
&=\sum_{a\in \mathcal{A}}\sum_{\nu_1, \nu_2\in\bohr} \alpha_{-\nu_2, -\nu_1} ((A^{a\dagger})_{-\nu_1})^\dagger (\bigcdot) (A^{a\dagger})_{-\nu_2} & \text{(Since if $A^a\in \mathcal{A}$ then $A^{a\dagger}\in\mathcal{A}$)}\nonumber\\
&=\sum_{a\in \mathcal{A}}\sum_{\nu_1, \nu_2\in\bohr} \alpha_{-\nu_2, -\nu_1} A^a_{\nu_1}(\bigcdot)(A^a_{\nu_2})^\dagger & \text{(Since $(A_\nu)^\dagger=(A^\dagger)_{-\nu}$)}\label{eq:Tdagger}
\end{align}
Comparing \autoref{eq:Tconjugated} and \autoref{eq:Tdagger}, we see that in order for $\mathcal{T}$ to satisfy the detailed balance condition $\mathcal{T}^\dagger =  \Gamma_\rho^{-1}\circ \mathcal{T} \circ \Gamma_\rho$, we require that the kernel $\alpha_{\nu_1, \nu_2}$ satisfies the condition
\begin{equation}
\alpha_{-\nu_2, -\nu_1} = \alpha_{\nu_1, \nu_2}e^{\frac{\beta(\nu_1+\nu_2)}{2}}.\label{eq:alphaDB}
\end{equation}
We highlight the crucial role played  by the requirement that every jump operator in $\mathcal{T}$ is `paired' with its Hermitian conjugate. Without this, since the left/right action of $A$ and $A^\dagger$ are exchanged under the transformation from $\mathcal{T}$ to $\mathcal{T}^\dagger$, it would not be possible to relate $\mathcal{T}$ and $\mathcal{T}^\dagger$. Physically, the detailed balance parametrized by \autoref{eq:rhoDB} (or \autoref{eq:alphaDB}) relates `downhill' transitions driven by $A_\nu$ with `uphill' transitions driven by $(A^\dagger)_{\nu}$. Note that this is related to the observation (Appendix B of Ref.~\onlinecite{lloyd2024quasiparticle})
 that the rate equation for `quasiparticle cooling' is detailed-balanced only if the jump operators are chosen to be Hermitian. 

The terms $\mathcal{G}$ and $\mathcal{R}$ are of the forms $\mathcal{G}[\bigcdot]=-i[G,\bigcdot]$ and $\mathcal{R}[\bigcdot]=-\frac{1}{2}\{R,\bigcdot\}$, respectively.
We now define $K = G -i\frac{R}{2}$, so that $(\mathcal{G}+\mathcal{R})[\bigcdot] = -i K(\bigcdot) + i(\bigcdot)K^\dagger$, and $(\mathcal{G}+\mathcal{R})^\dagger[\bigcdot]  =
i K^\dagger(\bigcdot)  -i(\bigcdot)K$. 
We can then write
\begin{align}
(\mathcal{G}+\mathcal{R})^\dagger[\bigcdot] - [\Gamma_{\rho_\beta}^{-1} \circ (\mathcal{G}+\mathcal{R}) \circ \Gamma_{\rho_\beta}][\bigcdot] &= i\left(K^\dagger(\bigcdot)  -(\bigcdot)K \right) + i\left( \Lambda_{\rho_\beta}[K](\bigcdot) - (\bigcdot) \Lambda_{\rho_\beta}^{-1}[K]\right)\nonumber
\\&= i(K^\dagger +\Lambda_{\rho_\beta}[K])(\bigcdot) - i(\bigcdot) (K+ \Lambda_{\rho_\beta}[K]^\dagger),\label{eq:GRdb}
\end{align}
where we have used the fact that if $\mathcal{Q}[\bigcdot] = Q_1(\bigcdot) \pm (\bigcdot)Q_2$, then
\begin{equation}
\Gamma_{\rho}^{-1}\circ \mathcal{Q} \circ \Gamma_{\rho}[\bigcdot] = \rho^{-1/2}\left(Q_1 \left(\rho^{1/2}(\bigcdot)\rho^{1/2}\right) \pm \left(\rho^{1/2}(\bigcdot)\rho^{1/2}\right) Q_2 \right)\rho^{-1/2}
 =\Lambda_\rho[Q_1](\bigcdot) \pm (\bigcdot) \Lambda_\rho^{-1}[Q_2],
 \end{equation}
and that $\Lambda_\rho^{-1}[K^\dagger] = \Lambda_\rho^{-1}[G] + \frac{i}{2}\Lambda_\rho^{-1}[R] = \Lambda_\rho[G]^\dagger + \frac{i}{2}\Lambda_\rho[R]^\dagger =\left(\Lambda_\rho[G] -\frac{i}{2}\Lambda_\rho[R]\right)^\dagger = \Lambda_\rho[K]^\dagger$.

Since $\mathcal{G}+\mathcal{R}$ must satisfy detailed balance \autoref{eq:rhoDB} independently of $\mathcal{T}$, we require that the RHS of  \autoref{eq:GRdb} vanishes. This in term requires that
\begin{equation}
    K^\dagger + \Lambda_{\rho_\beta}[K] = 0 + \lambda I
\end{equation}
where $\lambda \in \mathbb{R}$. For simplicity, we will set $\lambda=0$. Now, since $G, R$ are Hermitian, we can write
\begin{align}
K^\dagger + \Lambda_{\rho_\beta}[K] = G  + \frac{i}{2} R +\Lambda_{\rho_\beta}[G] -\frac{i}{2}\Lambda_{\rho_\beta}[R].
\end{align}
Using the explicit forms of $G$ and $R$ in \autoref{eq:TRGdefs} and 
\begin{align}
    \Lambda_{\rho_\beta}[(A^a_{\nu_2})^\dagger A^a_{\nu_1}]  &= \rho_\beta^{-1/2}(A^a_{\nu_2})^\dagger A^a_{\nu_1} \rho_\beta^{1/2}\nonumber\\&= e^{\frac{\beta(\nu_1-\nu_2)}{2}}(A^a_{\nu_2})^\dagger A^a_{\nu_1},
\end{align}
we find that 
\begin{equation}
    K^\dagger + \Lambda_{\rho_\beta}[K] =\sum_{a\in \mathcal{A}}\sum_{\nu_1, \nu_2} \left[ \left( 1 +e^{\frac{\beta(\nu_1-\nu_2)}{2}} \right)g_{\nu_1, \nu_2} +\frac{i}{2}\left( 1 -e^{\frac{\beta(\nu_1-\nu_2)}{2}} \right)\alpha_{\nu_1, \nu_2} \right](A^a_{\nu_2})^\dagger A^a_{\nu_1} 
\end{equation}
Setting this to zero gives us a condition on kernels of the decay and coherent parts such that they together satisfy detailed balance:
\begin{align}
 g_{\nu_1, \nu_2} = \frac{i}{2} \frac{1 -e^{\frac{\beta(\nu_1-\nu_2)}{2}}}{1 +e^{\frac{\beta(\nu_1-\nu_2)}{2}}} \alpha_{\nu_1, \nu_2}= -\frac{1}{2i} \tanh\frac{\beta(\nu_1-\nu_2)}{4}\alpha_{\nu_1, \nu_2}.\label{eq:gexact} 
\end{align}

\section{Detailed Derivation of Error Bounds I: Lindbladian Evolution and Magnus Expansion\label{app:bound_magnus}}
 In this section, we provide a detailed derivation of the first of the two error bounds used in the main text to derive the upper bound on the total Hamiltonian evolution time needed to prepare a thermal state up so some error $\nB \epsilon$ in trace distance. 

Here, we bound the errors incurred by the Lindbladian evolution and the Magnus expansion: we show that given $\mathcal K$ as in \autoref{eq:channelev}, and with $\mathcal L_T$ given in \autoref{eq:LDS}, that
\begin{equation}
    \norm{e^{J^2 \mathcal L_{T}}[\rho] - \mathcal K[\rho]}_1 
    = O\left(J^4 \left(\nB \left(\frac{T}{\sigma}\right)^4+\nB^2\right)\e^{\frac{\beta^2}{2 \sigma^2}}\right) .
\end{equation}

\subsection{Bounding the Magnus Approximation Error}

Recall that in the interaction picture, we implement the Hamiltonian
\begin{equation}
		\tilde{H}_{\rm SB}(t) = \sum_a J \{f(t) B_a^\dagger(t) \otimes A_a(t) + f^*(t) B_a(t)\otimes A_a^\dagger(t)\}.\label{eq:Hint2}
	\end{equation}
	Here $B= \tfrac{1}{2}(X_B  - i Y_B)$, and $B^\dagger = \tfrac{1}{2}(X_B + i Y_B)$ are lowering and raising operators for the ancilla, and $f(t)$ is given by \autoref{eq:chosenfilterfunction}.
	
Using the Magnus expansion, we may express the exact time evolution operator $\tilde{V}$ (see \autoref{eq:Vexact}, reproduced here) for convenience
\begin{equation}
\tilde{V} = \mathcal{T}\exp\left(-i J \int_{-T/2}^{T/2} dt \tilde{H}_{\rm SB}(t) \right)
\end{equation}
up to third order as 
	\begin{align}\label{eq:Vmagnus2}
			\tilde{V} \approx e^{-i \sum_{n=1}^3\tilde{H}_M^{(n)}}
	\end{align}
	with
	\begin{align}
		\tilde{H}_M^{(1)} &= \int_{-{T}/{2}}^{T/{2}}d t_1\, \tilde{H}_{\rm SB}(t_1), \\\nonumber
		\tilde{H}_M^{(2)} &= \frac{1}{2i}\int_{-{T}/{2}}^{T/{2}}d t_1\int_{-{T}/{2}}^{t}d t_2\ [\tilde{H}_{\rm SB}(t_1), \tilde{H}_{\rm SB}(t_2)]\nonumber\\
		\tilde{H}_M^{(3)} &= -\frac{1}{6} \int_{-T/2}^{T/2} dt_1 \int_{-T/2}^{t_1} dt_2 \int_{-T/2}^{t_2} dt_3\,
		\Big( 
		[\tilde{H}_{\rm SB}(t_1), [\tilde{H}_{\rm SB}(t_2), \tilde{H}_{\rm SB}(t_3)]] 
		+ [\tilde{H}_{\rm SB}(t_3), [\tilde{H}_{\rm SB}(t_2), \tilde{H}_{\rm SB}(t_1)]]
		\Big)\label{eq:magnusterms}.
	\end{align}
	Note that the higher-order terms include commutators between different jump operators (and hence, given our construction, different ancillas $a$, $a'$).

    The error incurred can be bounded by using (a mild generalization of) Theorem 1 of Ref.~\onlinecite{sharma2024simulation}. Adapting their notation to the present setting,  Ref.~\onlinecite{sharma2024simulation}  considers the question of approximating the time evolution operator
    corresponding to evolution under a (potentially time-dependent) $k$-local Hamiltonian  $H_{\rm SB}$, each term of which is supported on a subset $X$ of at most $k$ qubits, in the interaction picture of a geometrically local Hamiltonian $H$. We then have 
    \begin{equation}
        \norm{\tilde{V} - \exp\left(-i\sum_{n=1}^q \tilde{H}_{M}^{(n)} \right)}_{\infty} = O(\nB (J d T)^{q+1})\label{eq:magnusboundrephrased}
    \end{equation}
where
    $d = \max_{i} \sum_{X: i\in X} \norm{H_{\rm SB}}_{\infty}$ is the {\it interaction degree} (where the maximum is taken over all sites in the system and bath) of $B_X$, and $\tilde{H}_{\rm SB}(t) = e^{i Ht} {H}_{\rm SB}e^{-i Ht}$ is the time-evolved interaction-picture operator. In our case, $X$ consists of a single site and its ancilla, and  $H_{\rm SB}(t) =\sum_a f(t) A^\dagger_a(t)\otimes B_a(t) +\rm{h.c.}$ is explicitly time-dependent. Note that Ref. ~\onlinecite{sharma2024simulation} explicitly considers time-independent $B_X$, but observe that the generalization to time-dependent $B_X$ is immediate given the nature of their proof as long as $d$ can be bounded at all times. A second comment is in order regarding the scaling with $\nB$ in \eqref{eq:magnusboundrephrased}; Ref.~\onlinecite{sharma2024simulation} do not have a separation of system and bath and so the operator that plays the role $\tilde{H}_{\rm SB}$ simply scales with the system size. Analysing the derivation of their bound (or equivalently, by applying it to each term in the sum within $\tilde{H}_{\rm SB}$ separately) it is clear that the appropriate scaling in our case is with $\nB$ rather than the total size of system and bath together, $\nS +\nB$. In practice of course for effective cooling, $\nB\propto \nS$ so the distinction is not fundamental, but may be relevant for practical considerations where prefactors matter.

    We see that given the choice of filter function, since $\norm{A^\dagger_a(t)\otimes B_a(t)}_{\infty}=1$, we have 
    \begin{equation}
    d =  \max_t |f(t)|=\sqrt{\frac{2}{\pi\sigma^2}}e^{\frac{\beta^2}{8 \sigma^2}}.
\end{equation}
Combining this with \autoref{eq:magnusboundrephrased}, we arrive at \autoref{eq:Vmagnus}.

\subsection{Bounding the Error of the Reset Protocol}

It remains to bound the error of the reset protocol, i.e. the fact that we implement a Lindbladian by effectively ``Trotterizing": the drive followed by reset results in a quantum channel that approximates evolution with the Lindbladian for a short time.

Recall that each time step of our evolution implements the quantum channel
\begin{equation}
\mathcal{K}[\rho] = \tr_B \left[ \tilde{V} \left(\rho^0_B\otimes \rho\right)\tilde{V}^\dagger  \right],
\end{equation}
with
$\rho^0_B =\bigotimes_{a=1}^{\nB} \ket{0_a}\bra{0_a}_B$. Consider $\mathcal{L}_T$ as defined in \autoref{eq:LDScombined}. We argue that if we take $J = \sqrt{\Delta \tau}$, then 
\begin{align}
			\norm{e^{\Delta\tau \mathcal{L}_{T}}[\rho]-\mathcal{K}[\rho]}_1= O\left(J^4 \left(\nB \left(\frac{T}{\sigma}\right)^4+\nB^2\right)\e^{\frac{\beta^2}{2 \sigma^2}}\right).
\end{align}
Note that the scaling is one order in $J$ better than naively expected; this is due to the trace over the bath degrees of freedom. We now derive this in detail.

We begin by defining the Magnus-approximated channel
		\begin{align}
			{\mathcal{K}}_M[\rho] = \tr_B \left[  \exp\left(-i\sum_{n=1}^q \tilde{H}_{M}^{(n)} \right) \left(\rho^0_B\otimes \rho\right) \exp\left(i\sum_{n=1}^q \tilde{H}_{M}^{(n)} \right) \right]
		\end{align}
Using the triangle inequality,
\begin{align}
				\norm{e^{\Delta\tau \mathcal{L}_{T}}[\rho]-\mathcal{K}[\rho]}_1\leq 	
                \norm{e^{\Delta\tau \mathcal{L}_{T}}\rho-{\mathcal{K}}_M[\rho]}_1+	\norm{{\mathcal{K}}_M[\rho]-\mathcal{K}[\rho]}_1
		\end{align}
		The second term has the scaling, using a triangle and Hoelder inequality
		
        \begin{multline}
			\norm{{\mathcal{K}}_M[\rho]-\mathcal{K}[\rho] }_{1}=\norm{\tr_B \left(\tilde{V}-e^{-i \sum_{n=1}^3\tilde{H}_M^{(n)}}\right) \rho
            \left(\tilde{V}^\dagger-e^{i \sum_{n=1}^3\tilde{H}_M^{(n)}}\right)}_1
            \\\leq \left(\norm{\tilde{V}^\dagger}_\infty+\norm{e^{i \sum_{n=1}^3\tilde{H}_M^{(n)}}}_\infty\right)\norm{\rho}_1 \norm{\tilde{V}-e^{-i \sum_{n=1}^3\tilde{H}_M^{(n)}}}_\infty = \norm{\rho}_1 \, O\left(\nB J^4 \left(\frac{T}{\sigma}\right)^4 \e^{\frac{\beta^2}{2\sigma^2}}\right) ,
		\end{multline}
        
		due to the Magnus expansion error bound \autoref{eq:Vmagnus2} combined with the fact that the partial trace  (over the ancillas) is a contractive operation.
        
It remains to bound the first term. To do so, we show that $e^{\Delta\tau \mathcal{L}_{T}}\rho$ and ${\mathcal{K}}_M[\rho]$ agree up to $O(\nB J^4)$.
We first bound the induced trace norm of $\cL_{T}$:
We obtain for the transition term, using Hoelder's inequality
\begin{align}\label{eq:LTboundtransHoelder}
     \norm{L_{a;T}\rho L_{a;T}^\dagger}_1\leq \norm{L_{a;T}}_\infty^2 \norm{\rho}_1
\end{align}
and
\begin{align}
    \norm{L_{a;T}}_\infty\leq \int_{-T/2}^{T/2} |f(t)| \norm{A_a(t)}_\infty dt\leq \int_{-\infty}^{\infty} |f(t)|=\e^{\frac{\beta^2}{8\sigma^2}}.
\end{align}
Similarly, the dissipative and coherent parts can be bound, which gives
\begin{align}\label{eq:LTboundtrans}
    \norm{\cL_T}_{1\rightarrow 1}\leq 3 n_B\e^{\frac{\beta^2}{4\sigma^2}}
\end{align}
With the identification $J^2=\Delta \tau$ and a Taylor expansion, this gives  
	\begin{align}\label{eq:tLindblad}
		\norm{e^{\Delta \tau \mathcal{L}_{T}}[\rho]-(\rho+\Delta\tau \mathcal{L}_{T}[\rho])}_1= O(J^4 \norm{\cL_T}_{1\rightarrow 1}^2)=O(J^4 n_B^2\e^{\frac{\beta^2}{2\sigma^2}}),
	\end{align}  

On the other hand, we consider the expansion of ${\mathcal{K}}_M[\rho]$.
	Due to the trace over ancilla qubits at the end of the protocol and the structure of the system-bath interactions, all odd orders in $J$ cancel. It follows that
	\begin{align}\label{eq:defKrho}
	\begin{split}
		{\mathcal{K}}_M[\rho] &=  \rho+\tr_B \left[\tilde{H}_M^{(1)} \left(\rho^0_B\otimes \rho\right) \tilde{H}_M^{(1)} \right]-\frac{1}{2}\tr_B \left[\left(\tilde{H}_M^{(1)}\right)^2 \left(\rho^0_B\otimes \rho\right)+ \left(\rho^0_B\otimes \rho\right)\left(\tilde{H}_M^{(1)}\right)^2 \right]\\&
		-i \tr_B \left[\tilde{H}_M^{(2)} \left(\rho^0_B\otimes \rho\right)- \left(\rho^0_B\otimes \rho\right)\tilde{H}_M^{(2)} \right]
		+O\left(\nB^2 J^4 \e^{\frac{\beta^2}{2\sigma^2}}\right).
	\end{split}
	\end{align}
    Due to the tracing out of bath degrees of freedom, all jump operators have to appear in pairs and we obtain a more favorable scaling with $n_B^2$ instead of $n_B^4$.
    
	We consider the first term in detail; the treatment of the other terms proceeds analogously. We have  
	\begin{align}
	\begin{split}
		&\tr_B \left[\tilde{H}_M^{(1)} \left(\rho^0_B\otimes \rho\right) \tilde{H}_M^{(1)} \right]\\&=J^2 \sum_{a,a'} \int_{-{T}/{2}}^{T/{2}}d t_1 \int_{-{T}/{2}}^{T/2}d t_2
		 \tr_B\left[\left(f(t_1) B_a^\dagger(t_1) \otimes A_a(t_1) + f^*(t_1) B_a(t_1)\otimes A_a^\dagger(t_1)\right)
		 \left(\rho^0_B\otimes \rho\right)\right.
		 \\&\phantom{\sum_{a,a'} \int_{-{T}/{2}}^{T/{2}}d t_1 \int_{-{T}/{2}}^{T/2}d t_2} 
		 \left.\left(f(t_2) B_{a'}^\dagger(t_2) \otimes A_{a'}(t_2) + f^*(t_2) B_{a'}(t_2)\otimes A_{a'}^\dagger(t_2)\right)
		 \right]\\
	\end{split}
	\end{align}
	Tracing out the bath degrees of freedom forces $a=a'$. Meanwhile, placing the bath degree of freedom in the $\ket{0}$ state (the reset step) forces the application of $B_a^\dagger(t_1)$ on the left and $B_a(t_2)$ on the right.
Together, these give
	\begin{align}
			\tr_B \left[\tilde{H}_M^{(1)} \left(\rho^0_B\otimes \rho\right) \tilde{H}_M^{(1)} \right]	 = J^2 \sum_a \int_{-{T}/{2}}^{T/{2}}d t_1 \int_{-{T}/{2}}^{T/2}d t_2
		\,f(t_1) f^*(t_2) A_a(t_1) \rho A_a^\dagger(t_2).
	\end{align} 
	Identifying $\Delta\tau=J^2$, this gives the transition part in \autoref{eq:LDScombined}. Similarly, the remaining terms in the expansion \autoref{eq:defKrho} of ${\mathcal{K}}_M[\rho]$ can be respectively identified with the decay and coherent parts of $\mathcal{L}_T$, completing the proof.

\section{Detailed Derivation of Error Bounds II: Fixed-Point Error\label{app:boundfpe}}

In this section, we consider the second of the two error bounds used in the main text. Namely, we obtain an upper bound on the  overall fixed-point error, i.e. the difference between the fixed point of the local driving sampler $\mathcal L_{T}$ in \autoref{eq:LDS}, (that we denote $\rho_T$), and the thermal state $\rho_{\beta}$.

We split the total error into two parts using the triangle inequality, viz.
\begin{equation}
    \norm{\rho_T  - \rho_\beta}_1 \leq \norm{\rho_T - \rho_{\infty}}_1 + \norm{\rho_{\infty} - \rho_{\beta}}_1
\end{equation}
where $\rho_\infty$ is the fixed point of  $\mathcal{L}\equiv\lim_{T\to\infty}\mathcal{L}_T$, the infinite time evolution limit of the local driving sampler, i.e.  $\mathcal{L}[\rho_\infty]=0$. 
The two terms on the RHS of the above equation capture different pieces of physics. The first is the error incurred by the fact that the exact sampler involves an infinite-time smoothing, while the local driving sampler is truncated at time $T$. The the second essentially comes from the fact that even as $T\to \infty$, the local driving sampler has the ``wrong'' coherent part, given by the Lamb shift, which we tackle via the so-called ``secular approximation''. We address these in turn.

Note that throughout this section to streamline notation we consider a single jump operator $A$; the generalization to multiple jump operators $A_a$ is straightforward and simply involves a sum over $a$, and all the errors simply pick up  an overall factor of $\nB = |\mathcal{A}|$.

\subsection{Error from finite-time evolution\label{app:bound_finite_time}}
Here, we prove that
\begin{align}
    \norm{\rho_T - \rho_{\infty}}_1 
    \leq 4\,\tmix\, \norm{\mathcal L_{T} - \mathcal L_{\infty}}_{1-1}
    \leq  \frac{24 \sqrt{2}}{\sqrt{\pi}} 
       \left(\frac{\sigma}{T}\right) \e^{\frac{\beta^2}{4\sigma^2}} \e^{-\frac{T^2}{2\sigma^2}} t_{\text{mix}}(\mathcal{L}_T)
\end{align}
We can use Lemma II.1 of \cite{CKBG23} to write 
	\begin{align}
		\norm{\rho_T-\rho}_1\leq 4 \,t_{\text{mix}}(\mathcal{L}_T) \norm{\mathcal{L}_{T}-	\mathcal{L}_\infty}_{1-1}.
	\end{align}
Recall that
\begin{equation}
		 \mathcal{L}_{T}[\cdot] = \int_{-{T}/{2}}^{T/{2}}d t_1 \int_{-{T}/{2}}^{T/2}d t_2\,f(t_1) f^*(t_2)\left( -i \left[-\frac{\text{sgn}(t_1-t_2)}{2i}A^\dagger(t_2) A(t_1), \cdot \right]  + A(t_1) (\cdot) A^\dagger(t_2) -\frac{1}{2} \left\{A^\dagger(t_2) A(t_1), \cdot\right\} \right).
	\end{equation}
Recall that $\mathcal{L}_\infty =\lim_{T\to\infty} \mathcal{L}_T$, and consider the transition parts of the Lindbladians: We obtain, using Hoelder's and the triangle inequality and $\|A(t)\|_\infty\leq 1$:
\begin{multline}
\norm{L_{a;T}\rho L_{a;T}^\dagger-L_{a;\infty}\rho L_{a;\infty}^\dagger}_1=\norm{(L_{a;T}-L_{a;\infty})\rho L_{a;T}^\dagger+ L_{a;\infty}\rho (L^\dagger_{a;T}-L^\dagger_{a;\infty})}_1\\\leq \norm{\rho}_1 \norm{L_{a;T}-L_{a;\infty}}_\infty \left(\norm{L_{a;T}}_\infty+\norm{L_{a;\infty}}_\infty\right)
\\\leq 2 \norm{\rho}_1 \, \left(\int_{-\infty}^\infty\rd t |f(t)|\right)\left(\int_{\mathbb{R}_{[T]}}\rd t |f(t)|\right), 
\end{multline}
where we have defined $\mathbb{R}_{[T]}= \mathbb{R}\backslash[-T/2,T/2]$. Similarly, we can bound the dissipative part and the coherent part, using
$|\frac{\text{sgn}(t_1-t_2)}{2}|<\frac{1}{2}$.
Since this applies for all $\rho$, we obtain for the induced norm
  
\begin{align}
		\norm{\mathcal{L}_T-	\mathcal{L}_\infty}_{1-1}\leq 6 \left(\int_{-\infty}^\infty\rd t |f(t)|\right)\left(\int_{\mathbb{R}_{[T]}}\rd t |f(t)|\right)
\end{align}
 For our specific choice of filter function 
	\begin{align}
		f(t)=\sqrt{\frac{2}{\pi \sigma^2}}\e^{-\frac{2}{\sigma^2}\left(t-\frac{\I \beta}{4}\right)^2},
	\end{align}
    we obtain 
	\begin{align}
		\int_{\mathbb{R}_{[T]}}\rd t_1 |f(t_1)|=\e^{\frac{\beta^2}{8\sigma^2}}\mathrm{erfc}\left( \frac{T}{\sigma \sqrt{2}} \right)\leq \frac{\sqrt{2}\sigma}{\sqrt{\pi}T}\e^{\frac{\beta^2}{8\sigma^2}}\e^{-\frac{T^2}{2\sigma^2}},
	\end{align}
	where the last bound holds for $\frac{T^2}{2\sigma^2}>1$, which is guaranteed by the choice in \autoref{eq:LDS_params} as $\epsilon\to 0$.
	This leads to the final bound
		\begin{align}
        \norm{\rho_T -\rho_\infty}_1 \leq  \frac{24 \sqrt{2}}{\sqrt{\pi}} 
       \left(\frac{\sigma}{T}\right) \e^{\frac{\beta^2}{4\sigma^2}} \e^{-\frac{T^2}{2\sigma^2}} t_{\text{mix}}(\mathcal{L}_T). 
	\end{align}

\subsection{Secular Approximation for the Lamb Shift\label{app:bound_secular}}
We next derive the inequality
\begin{equation}\label{eq:app:bound_secular_final}
    \norm{\rho_{\infty} - \rho_{\beta}}_1 = \Tilde O\left(
        \nB\, \frac{\beta}{\sigma}\, \max(\tmix(\mathcal L_{\infty}), \tmix(\mathcal L_{\rm diss})
    \right)
\end{equation}
The derivation ultimately proceeds by means of implementing the so-called ``secular approximation'', but first we need to perform  some preliminary manipulations involving eigenvector perturbation theory. Much of our treatment is agnostic to specific choices of the filter function, except the very last portions of the proof which impose some requirements of the time-domain asymptotics of $f(t)$. Accordingly we will only specify our choice of $f(t)$ where necessary.

\subsubsection{Preliminaries.} 
We first establish some notation. We denote by $\mathcal{L}_\infty, \mathcal{L}_{\rm sec}$, and $\mathcal{L}_\beta$ the Lindbladians that respectively correspond to the $T\to \infty$ limit of the local-driving sampler $\mathcal{L}_T$, its {\it secular approximation} (to be explained below), and the exact sampler.  These all have identical dissipative parts, but differ in their coherent parts:
\begin{align}\label{eq:Ldefs}
\mathcal{L}_\beta[\bigcdot] &= -i[G, \bigcdot ] + \mathcal{L}_{\rm diss}[\bigcdot]\nonumber\\
\mathcal{L}_\infty[\bigcdot] &= -i[H_{\rm LS}, \bigcdot ] + \mathcal{L}_{\rm diss}[\bigcdot]= -i [B, \bigcdot] + \mathcal{L}_\beta[\bigcdot]  \nonumber\\
\mathcal{L}_{\rm sec}[\bigcdot] &= -i[B_{\rm sec}, \bigcdot] + \mathcal{L}_\beta[\bigcdot] =  -i[B_{\rm sec} +G, \bigcdot] + \mathcal{L}_{\rm diss}[\bigcdot]
\end{align}
where we have defined $B =  H_{\rm LS} - G$, the difference between the coherent parts of the local-driving and exact samplers, and its secular approximation $B_{\rm sec}$ (to be specified below). Note that in contrast to Ref. \onlinecite{CKBG23} (Appendix D), we {\it do not} make a secular approximation for the dissipative part of the sampler. This is because we are bounding the difference between the local-driving sampler and the {\it exact} Gibbs sampler, whereas Ref. \onlinecite{CKBG23} were bounding the difference between a physical system-bath problem and a suitably smoothed approximation of a Davies Lindbladian {\it without} the coherent part necessary to make the latter an exact Gibbs sampler. This difference considerably streamlines our proof relative to that of Ref. \onlinecite{CKBG23}. 

The Lindbladians introduced above satisfy the following fixed-point relations:
\begin{equation}
    \mathcal{L}_\beta[\rho_\beta] = 0; \quad\quad\mathcal{L}_\infty[\rho_\infty] =0; \quad\quad \mathcal{L}_{\rm sec}[\rho_{\rm sec}] =0,
\end{equation}
where the final equation can be viewed as a {\it definition} of $\rho_{\rm sec}$.

We now have by the triangle inequality that
\begin{equation}\label{eq:normtriangle}
    \norm{\rho_\infty-\rho_\beta}_{1} \leq \norm{\rho_\infty-\rho_{\rm sec}}_{1} + \norm{\rho_{\rm sec} -\rho_\beta}_{1}
\end{equation}

Of the remaining two expressions in \autoref{eq:normtriangle}, $\norm{\rho_\infty-\rho_{\rm sec}}_{1}$ can be bounded using Lemma II.1 of Ref. \onlinecite{CKBG23}, yielding
\begin{equation}
    \norm{\rho_\infty-\rho_{\rm sec}}_{1}  \leq 4\norm{\mathcal{L}_\infty - \mathcal{L}_{\rm sec}}_{1-1}\cdot t_{\rm mix}(\mathcal{L}_\infty).
\end{equation}
As we will frequently do, we can use H\"older's inequality to relate $p-p$ norms of superoperators $\mathcal{Q}[\cdot] = -i[Q, \cdot]_{\pm} $ whose sole action is commutation ($-$) or anticommutation ($+$) with some operator $\mathcal{Q}$ to the operator norm of $\mathcal{Q}$, yielding \footnote{This follows since $\norm{\mathcal{Q}}_{p-p} \equiv \sup_{X\neq 0} \frac{\norm{\mathcal{Q}[X]}_{p}}{\norm{X}_p} = \frac{\norm{-i[{Q},X]_{\pm}}_{p}}{\norm{X}_p} \leq \sup_{X\neq 0} \frac{2\norm{\mathcal{Q}}_\infty\norm{X}_p}{\norm{X}_p} = \norm{Q}_\infty$  where we used the triangle inequality to rewrite the commutator in terms of a product and then applied H\"older's inequality to rewrite the numerator.}
\begin{equation}\label{eq:normrhorhosecBBsec}
    \norm{\rho_\infty-\rho_{\rm sec}}_{1}  \leq 8\norm{B - B_{\rm sec}}_\infty\cdot t_{\rm mix}(\mathcal{L}_\infty).
\end{equation}

The second piece is trickier to bound, and requires a use of eigenvector perturbation theory. Specifically, we use the following identity from Ref. \onlinecite{CKBG23}, Appendix E: given two linear [super]operators\footnote{The original argument in Ref.~\onlinecite{CKBG23} is in terms of matrices, but we can view these as superoperators written in the doubled space where density matrices are vectorized, so we give the argument in terms of superoperators  here for brevity.}$\mathcal{M}$, $\mathcal{M}'$ (such that $\mathcal{M}'$ is a `small' perturbation of $\mathcal{M}$), with eigenvectors $\rho$, $\rho'$ corresponding to eigenvalues $\lambda, \lambda'$, we have
\begin{equation}\label{eq:eigenvectorPT}
 \norm{\rho' - \rho}_2 \leq \frac{2\sqrt{2} \norm{\mathcal{M}' -\mathcal{M}}_{2-2} + |\lambda'-\lambda|}{\zeta_{-2}(\mathcal{M} - \lambda I)},
\end{equation}

where $\zeta_{-2}(\mathcal{Q})$ is the second smallest singular value of $\mathcal{Q}$.

Although in general eigenvector perturbation theory is poorly behaved for generic non-Hermitian operators (such as $\mathcal{M}, \mathcal{M}'$),  it is possible to choose both to share a common eigenvalue $\lambda=\lambda'=0$: simply choose superoperators $\mathcal{M}, \mathcal{M}'$ which have $\rho_\beta^{-1/4}\rho_{\rm sec}\rho_\beta^{-1/4}$ and $\rho_{\beta}^{1/2}$ as fixed points (eigenstates with eigenvalue 0).
It follows then
\begin{align}\label{eq:1normto2norm}
    \norm{\rho_\beta-\rho_{\rm sec}}_1&=
    \norm{\rho_{\beta}^{1/4}(\rho_\beta^{1/2}-\rho_\beta^{-1/4}\rho_{\rm sec}\rho_\beta^{-1/4})\rho_\beta^{1/4}}_1\nonumber\\
    &\leq \norm{\rho_\beta^{1/4}}_4^2\norm{\rho_\beta^{1/2}-\rho_\beta^{-1/4}\rho_{\rm sec}\rho_\beta^{-1/4}}_2\nonumber\\
    &= \norm{\rho_\beta^{1/2}-\rho_\beta^{-1/4}\rho_{\rm sec}\rho_\beta^{-1/4}}_2,  
\end{align}
where we used Hoelder's inequality in the second line and $\norm{\rho_\beta^{1/4}}_4=1$ in the last step. The remaining term can then be bounded using Eq.~\eqref{eq:eigenvectorPT}.

We now choose $\mathcal{M}$ as
\begin{align}
    \mathcal{M}[\bigcdot] =\rho_\beta^{-1/4}\mathcal{L}_{\rm sec}[\rho_\beta^{1/4}\bigcdot \rho_\beta^{1/4}]\rho_\beta^{-1/4}
\end{align}

However, there is a wide range of possible superoperators that can be chosen to have $\rho_\beta^{1/2}$ as a fixed point, due to the latter's central role in defining detailed balance. The nontrivial idea introduced in Ref. \onlinecite{CKBG23} is that this flexibility can be leveraged to obtain a tight bound. An elegant choice is
\begin{equation}
    \mathcal{M}'[\bigcdot]= \rho_\beta^{1/4}\mathcal{L}_2^\dagger[\rho_\beta^{-1/4}\bigcdot \rho_\beta^{-1/4}]\rho_\beta^{1/4},
\end{equation}
with  $\mathcal{L}_2^\dagger[\bigcdot] = -i[B_{\rm sec},\bigcdot] + \mathcal{L}_\beta^\dagger[\bigcdot]$. Observe that $\mathcal{M}'[\rho_\beta^{1/2}]  = \rho_\beta^{1/4}\mathcal{L}_2^\dagger[\mathds{1}]\rho_\beta^{1/4} =0$, by using the fact that 
$\mathcal{L}_2^\dagger[\mathds{1}] = 0$ as long as $\mathcal{L}_2$ is a valid Lindbladian. 

We then have, using the definition \autoref{eq:Ldefs} of $\mathcal{L}_{\rm sec}$ and by rewriting the detailed balance condition as
\begin{equation}
\rho_{\beta}^{-1/4}\mathcal{L}_\beta[\rho_\beta^{1/4}\bigcdot\rho_\beta^{1/4}]\rho_{\beta}^{-1/4} - \rho_\beta^{1/4}\mathcal{L}_\beta^\dagger[\rho_\beta^{-1/4}(\bigcdot) \rho_\beta^{-1/4}]\rho_\beta^{1/4}=0\nonumber
\end{equation}
that
\begin{align}
(\mathcal{M}' -\mathcal{M})[\bigcdot] &= \rho_\beta^{1/4}\mathcal{L}_2^\dagger[\rho_\beta^{-1/4}(\bigcdot) \rho_\beta^{-1/4}]\rho_\beta^{1/4} -\rho_\beta^{-1/4}\mathcal{L}_{\rm sec}[\rho_\beta^{1/4}(\bigcdot)\rho_\beta^{1/4}]\rho_\beta^{-1/4}\nonumber\\
& = -i\rho_\beta^{1/4}[B_{\rm sec},\rho_\beta^{-1/4} (\bigcdot) \rho_\beta^{-1/4}]\rho_\beta^{1/4} + \rho_\beta^{1/4}\mathcal{L}_\beta^\dagger[\rho_\beta^{-1/4}(\bigcdot) \rho_\beta^{-1/4}]\rho_\beta^{1/4} \nonumber \\&+ i\rho_\beta^{-1/4}[B_{\rm sec}, \rho_\beta^{1/4}(\bigcdot)\rho_\beta^{1/4}]\rho_\beta^{-1/4}  -\rho_\beta^{-1/4}\mathcal{L}_\beta[\rho_\beta^{1/4}(\bigcdot)\rho_\beta^{1/4}]\rho_\beta^{-1/4}\nonumber\\
& = i \left[\left(\rho_\beta^{-1/4} B_{\rm sec}\rho_\beta^{1/4} -  \rho_\beta^{1/4}B_{\rm sec} \rho_\beta^{-1/4}\right) (\bigcdot) - (\bigcdot)\left(\rho_\beta^{1/4}B_{\rm sec}\rho_\beta^{-1/4} -  \rho_\beta^{-1/4}B_{\rm sec} \rho_\beta^{1/4}\right)  \right]
\end{align}
Using the triangle and H\"older inequalities again, we then have
\begin{align}\label{eq:MMpnorm}
\norm{\mathcal{M}' -\mathcal{M}}_{2-2} &\leq \norm{\rho_\beta^{-1/4}B_{\rm sec}\rho_\beta^{1/4} -  \rho_\beta^{1/4}B_{\rm sec} \rho_\beta^{-1/4}}_\infty + \norm{\rho_\beta^{1/4}B_{\rm sec}\rho_\beta^{-1/4} -  \rho_\beta^{-1/4}B_{\rm sec} \rho_\beta^{1/4}}_\infty \nonumber\\
&= 2 \norm{\rho_\beta^{-1/4}B_{\rm sec}\rho_\beta^{1/4} -  \rho_\beta^{1/4}B_{\rm sec} \rho_\beta^{-1/4}}_\infty,
\end{align}
where in the second line we have used $\norm{Q}_* = \norm{Q^\dagger}_*$, with ${Q}=\rho_\beta^{-1/4}B_{\rm sec}\rho_\beta^{1/4} -  \rho_\beta^{1/4}B_{\rm sec} \rho_\beta^{-1/4}$.

Using the Fan-Hoffmann inequality \cite[Proposition III.5.1]{Bhatia1997book} on the second-smallest singular value\footnote{Note that for any $n\times n$ matrix $A$, $\zeta_j(A) = \zeta_j(-A) \geq \lambda_j\left(-\tfrac{1}{2}(A + A^\dagger)\right) = -\lambda_{n-j+1}\left(\tfrac{1}{2}(A + A^\dagger)\right)$, which yields the identity in \autoref{eq:FanHoffmann} if indexing is understood to be modulo $n$.} of $\mathcal{M}$ (not necessarily Hermitian) we have
\begin{equation}\label{eq:FanHoffmann}
\zeta_{-2}(\mathcal{M}  - \lambda I) \geq -\lambda_2\left(\frac{\mathcal{M}+\mathcal{M}^\dagger}{2} - {\rm Re}(\lambda) I\right)
= -\lambda_2\left({\mathcal{L}_{\beta}} - {\rm Re}(\lambda) I\right)
\end{equation}
where $\lambda_2(\mathcal{Q})$ denotes the second largest eigenvalue of $\mathcal{Q}$. Note that in the final step we have simply used  \autoref{eq:Ldefs} with $\frac{\mathcal{M}+\mathcal{M}^\dagger}{2} = \rho_\beta^{-1/4}\mathcal{L}_{\beta}.[\rho_\beta^{1/4}\bigcdot\rho_\beta^{1/4}]\rho_\beta^{-1/4}$, which has the same spectrum as $\mathcal{L}_\beta$.
The spectral gap can be lower-bounded by the mixing time via
\begin{equation}\label{eq:l2mixingtimelink}
-\lambda_2(\mathcal{L}_{\rm beta}) \geq \frac{\ln(2)}{t_{\rm mix}(\mathcal{L}_{\beta})}.
\end{equation}

We can now assemble a bound on the second term in \autoref{eq:normtriangle}, by using
\autoref{eq:eigenvectorPT} and \autoref{eq:1normto2norm} with $\lambda=\lambda'=0$ and the results in \autoref{eq:MMpnorm}, \autoref{eq:FanHoffmann}, and \autoref{eq:l2mixingtimelink}:
\begin{equation}\label{eq:normrhosecrhobeta}
\norm{\rho_{\rm sec} -\rho_\beta}_{1} \leq \frac{1}{\ln 2}\norm{\rho_\beta^{-1/4}B_{\rm sec}\rho_\beta^{1/4} -  \rho_\beta^{1/4}B_{\rm sec} \rho_\beta^{-1/4}}_\infty\cdot t_{\rm mix}(\mathcal{L}_{\beta}) 
\end{equation}

Putting together \autoref{eq:normrhorhosecBBsec} and \autoref{eq:normrhosecrhobeta}, we finally arrive at a bound on the error of the fixed point of local-driving sampler in terms of secular-approximable quantities and mixing times

\begin{align}\label{eq:errorintermsofsecnorms}
 \norm{\rho_{\infty} - \rho_\beta}_1 &\leq   8\norm{B - B_{\rm sec}}_\infty\cdot t_{\rm mix}(\mathcal{L}_\infty) + \frac{1 }{\ln 2}\norm{\rho_\beta^{-1/4}B_{\rm sec}\rho_\beta^{1/4} -  \rho_\beta^{1/4}B_{\rm sec} \rho_\beta^{-1/4}}_\infty \cdot t_{\rm mix}(\mathcal{L}_{\beta})\nonumber\\
  &\leq 8\left(\norm{B - B_{\rm sec} }_\infty+ \norm{\rho_\beta^{-1/4}B_{\rm sec}\rho_\beta^{1/4} -  \rho_\beta^{1/4}B_{\rm sec} \rho_\beta^{-1/4}}_\infty\right)\cdot \max(t_{\rm mix}(\mathcal{L}_\infty), t_{\rm mix}(\mathcal{L}_{\beta})).
\end{align}

We now turn to computing these norms using the secular approximation, which will allow us to bound the terms in the  RHS of \autoref{eq:errorintermsofsecnorms}.

\subsubsection{Implementing the Secular Approximation.} 
The secular approximation is most transparently phrased in frequency space (although bounding various contributions is often easier in the time domain). Consider a generic  operator of the form  $B = \sum_{\nu_1,\nu_2\in \bohr} b_{\nu_1, \nu_2} (A_{\nu_2})^\dagger A_{\nu_1}$. The secular approximation of $B$ is obtained by restricting the sum to the ``almost diagonal'' pieces, such that $|\nu_1-\nu_2|\lesssim \mu$, where $\mu$ is some suitably chosen cutoff. Formally, we do this by multiplying the kernel $b_{\nu_1, \nu_2}$ by a suitably chosen ``bump function'',

\begin{equation}
B = \sum_{\nu_1,\nu_2\in \bohr} b_{\nu_1, \nu_2} (A_{\nu_2})^\dagger A_{\nu_1} \implies  B_{{\rm sec}} = \sum_{\nu_1,\nu_2\in \bohr} b_{\nu_1, \nu_2}w\left(\frac{\nu_-}{\mu}\right) (A_{\nu_2})^\dagger A_{\nu_1}
\end{equation}
where $\nu_- \equiv \nu_1 -\nu_2$ and $w(x)$ is a smooth (i.e., infinitely differentiable) function with
\begin{align}
	w(x)=\begin{cases}
		1,\qquad x=0\\
		0,\qquad|x|>1\\
		<1,\qquad \text{else}
		\end{cases}.
\end{align}
For our purposes, it is convenient to choose a bump function such that it remains close to $1$ except in some finite interval near the boundaries at $\pm1$. While a specific choice will not be essential to our argument (modulo this requirement) a concrete  choice is
\begin{equation}
	w(x)=w_1\left(\frac{x+1}{1-\lambda}\right)w_1\left(\frac{1-x}{1-\lambda}\right).
\end{equation}
where we define $w_1(x)$ in terms of another auxiliary function
\begin{equation}
		w_1(x)=\frac{w_2(x)}{w_2(x)+w_2(1-x)}, \quad \quad
		w_2(x)=\begin{cases}
			\exp\left(-\frac{1}{x}\right),\qquad &x>0\\
			0,\qquad  &\text{else}.
			\end{cases}
\end{equation}  
Note that $w$ is zero outside of $[-1,1]$ and one in the interval $[-\lambda,\lambda]$. By choosing $\lambda$ close to one, we can make the function arbitrarily sharp.

The  reason to implement the secular approximation  is that in the limit where we take $\mu\to 0$,  the secular piece is purely diagonal in Bohr frequencies and hence commutes with $\rho_\beta$. An important point is that, depending on the specific function we wish to approximate, a sharp cutoff in frequencies may prove difficult to bound; implementing the secular approximation then requires a judicious use of bump functions (smooth functions that are strictly vanishing outside a compact domain) in order to remove high-frequency terms. This will be the case for our coherent parts.

Evidently, in order to use the secular approximation we must bound  the two separate terms in \autoref{eq:errorintermsofsecnorms}: the  ``almost-commuting'' piece  $\rho_\beta^{-1/4}B_{\rm sec}\rho_\beta^{1/4} -  \rho_\beta^{1/4}B_{\rm sec} \rho_\beta^{-1/4}$ and the  non-secular piece $B- B_{\rm sec}$. The strategy of proving these bounds is distinct, so we tackle them in turn. 
\subsubsection{Bounding the Almost-Commuting Part.} First, we consider the `almost commuting' piece $\norm{\rho_\beta^{-1/4}B_{\rm sec}\rho_\beta^{1/4} -  \rho_\beta^{1/4}B_{\rm sec} \rho_\beta^{-1/4}}_\infty$.	
Note that we can always choose to write any operator in terms of its Bohr frequency expansion as $B = \sum_{\nu\in \bohr} B_\nu$.  Using the secular approximation, it then follows that
    \begin{align}
    	\norm{\rho_\beta^{-1/4}B_{\rm sec}\rho_\beta^{1/4} -  \rho_\beta^{1/4}B_{\rm sec} \rho_\beta^{-1/4}}_\infty=\norm{\sum_{\nu\in\bohr} w\left(\frac{\nu}{\mu}\right) \left(e^{\frac{\beta \nu}{4}}-e^{-\frac{\beta \nu}{4}}\right) B_\nu}_\infty.
    \end{align}
	We now exploit the fact that $w\left(\frac{\nu}{\mu}\right) \left(e^{\frac{\beta \nu}{4}}-e^{-\frac{\beta \nu}{4}}\right)<\beta \mu$ for $\beta \mu\leq1$, which places a restriction on the choice of $\mu$. From this, it follows that
	    \begin{align}
		\norm{\rho_\beta^{-1/4}B_{\rm sec}\rho_\beta^{1/4} -  \rho_\beta^{1/4}B_{\rm sec} \rho_\beta^{-1/4}}_\infty= O\left(\norm{\sum_{\nu\in\bohr} B_\nu}_\infty \beta \mu\right) =O\left(\norm{B}_\infty \beta \mu\right).
	\end{align}
Since $B = H_{\rm LS}-G$, the final step in bounding the almost-commuting piece is therefore to bound the norms of the operators that appear
in the coherent evolution of the exact sampler and the local driving sampler (the latter being the Lamb shift), which we may reduce to bounding the norm of each individually. For both cases, we can show that this is $O(1)$,  by writing the relevant operator as a  time domain integral of an expression quadratic in the jump operators multiplied by some weight function that depends quadratically on the smoothing functions. Since the jump operators can always be fixed to have unit norm by a suitable choice of normalization, we can use triangle inequalities to simplify the bound to one in terms of time domain integrals over the weight functions. For the  Lamb shift the time domain representation is shown in \autoref{eq:HLStimedomain}; the weight functions are just a product of filter functions and so resulting bound on the norm is clearly $O(1)$ due to the normalization of the filter functions. For the exact coherent part, the introduction of the $\tanh\frac{\beta\nu_-}{4}$ complicates the weight functions. It is straightforward but tedious to show 
[cf. Ref.~\onlinecite{CKG23}, Appendix A]  that
		\begin{align}\label{eq:coherenttime}
			G=
			\int_{-\infty}^{\infty}\!\tilde{g}_-(t_-)\e^{-\I H t_-} \left(\int_{-\infty}^{\infty}\tilde{g}_+(t_+)A(t_+)A(-t_+)\rd t_+ \right)\e^{\I H t_-}\rd t_-
		\end{align}
        	with
	\begin{align}
		\tilde{g}_-(t)=\frac{1}{2i} \int_{-\infty}^{\infty}d\nu_-\, e^{i\nu_-t_-}g_-(\nu_-)=\left[\frac{2\sqrt{2\pi}}{\beta \cosh\left(\frac{2\pi t_-}{\beta}\right)}\right]*_{t_-}\left[\frac{\sqrt{2}}{\sigma}e^{\frac{\beta^2-16t_-^2}{4\sigma^2}}\sin\left(\frac{2  \beta t_- }{\sigma^2}\right)\right]
	\end{align}
	and
	\begin{align}
		\tilde{g}_+(t)=\int_{-\infty}^{\infty}d \nu_+\,e^{i\nu_+t_+}g_+(\nu_+)=\frac{4 \sqrt{\pi} e^{-\frac{2}{\sigma^2}\left(t_+ - \frac{i\beta}{4}   \right)^2}}{\sigma}.
	\end{align}
With this in hand, we can use triangle inequalities to write
    	\begin{align}
		\norm{G}_\infty \leq \left(\int_{-\infty}^\infty {d}t_+ |\tilde{g}_+(t_+)|\right) \left(\int_{-\infty}^\infty {d}t_- |\tilde{g}_-(t_-)|\right).
	\end{align}
Since $g_+(t_+)$ is a shifted Gaussian, its integral is bounded. Now, $g_-(t_-)$ is the convolution of two bounded functions, and we can use Young's convolution inequality for the $L_1$ norm to bound the  norm of this convolution terms of the product of the norms. In each case due to the choice of overall normalization the bound is $O(1)$.  Combining the result above, we find that for a single jump operator, 
\begin{equation}
    \norm{\rho_\beta^{-1/4}B_{\rm sec}\rho_\beta^{1/4} -  \rho_\beta^{1/4}B_{\rm sec} \rho_\beta^{-1/4}}_\infty = O(\beta\mu)
\end{equation}
Finally, for multiple jump operators $\nB = |\mathcal{A}|$ the errors add independently, so that we finally have
\begin{equation}\label{eq:bound_almost_commuting}
    \norm{\rho_\beta^{-1/4}B_{\rm sec}\rho_\beta^{1/4} - \rho_\beta^{1/4}B_{\rm sec} \rho_\beta^{-1/4}}_\infty = O(\nB\beta\mu).
\end{equation}
\subsubsection{Bounding the Non-Secular Part.}
The operator $B$ that appears in the error bounds involves the difference between the coherent parts of the local driving sampler (the Lamb shift) and the exact sampler. A sufficient bound obtains by simply using the triangle inequality to bound these pieces separately. Therefore will proceed initially being agnostic to the form of $B$, making general arguments as to the form of the kernel such that the non-secular part $B - B_{\rm sec}$ is bounded, before we specialize to the form dictated by our choice of filter functions.

Consider the non-secular piece
\begin{equation}
B - B_{\rm sec} =  \sum_{\nu_1,\nu_2\in \bohr} b_{\nu_1, \nu_2}\left (1 -w\left(\frac{\nu_-}{\mu}\right)\right) (A_{\nu_2})^\dagger A_{\nu_1}
\end{equation}
Now, let us {\it define} a time-domain kernel $\mathcal{W}(t_1, t_2)$ implicitly via
\begin{equation} \label{eq:nonsecularkerneldef}
b_{\nu_1, \nu_2}\left (1 -w\left(\frac{\nu_-}{\mu}\right)\right) \equiv \int_{-\infty}^\infty dt_1 \int_{-\infty}^\infty dt_2\, \mathcal{W}(t_1, t_2) e^{i\nu_1 t_1 -i \nu_2 t_2}.
\end{equation}
Using \autoref{eq:nonsecularkerneldef} in our expression for the non-secular part, we have 
\begin{align}
B - B_{\rm sec} &= \sum_{\nu_1,\nu_2\in \bohr} \int_{-\infty}^\infty dt_1 \int_{-\infty}^\infty dt_2\, \mathcal{W}(t_1, t_2) e^{i\nu_1 t_1 -i \nu_2 t_2} (A_{\nu_2})^\dagger A_{\nu_1}\nonumber\\ &= \int_{-\infty}^\infty dt_1 \int_{-\infty}^\infty dt_2\, \mathcal{W}(t_1, t_2) A^\dagger(t_2) A(t_1)\label{eq:timedomainBnonsec}
\end{align}
where we have simply used the representation of time evolution in terms of Bohr frequencies. The reason for this step is that the  $A(t)$ are obtained by time-evolving bounded local operators, and therefore their norm has some $O(1)$ bound (which can be set to be $1$ by a suitable choice of normalization, which we henceforth assume). We then find, using \autoref{eq:timedomainBnonsec} and the triangle inequality, that the non-secular part can be bounded by a certain two-dimensional time domain integral:
\begin{equation}\label{eq:nonsecasTDnorm}
\norm{B - B_{\rm sec}}_\infty \leq \left| \int_{-\infty}^\infty dt_1 \int_{-\infty}^\infty dt_2\, \mathcal{W} (t_1, t_2) \right|.
\end{equation}
Now, we can invert \autoref{eq:nonsecularkerneldef} by recognizing that it is just a Fourier transformation, so that
\begin{equation}\label{eq:BasFT}
\mathcal{W}(t_1, t_2) = \int_{-\infty}^\infty \frac{d\nu_1}{2\pi}  \int_{-\infty}^\infty \frac{d\nu_2}{2\pi} b_{\nu_1, \nu_2}\left( 1- w\left(\frac{\nu_-}{\mu}\right)\right) e^{-i\nu_1 t_1 + i\nu_2 t_2}
\end{equation}

To proceed, we need to consider the form of $b_{\nu_1, \nu_2}$. For now, we will simply assume that this has the form $b_{\nu_1, \nu_2} = b_+({\nu_1 + \nu_2}) b_-({\nu_1 -\nu_2})$; this will be the form of the kernel for both $H_{{\rm LS}, f}$ and $G$. We then see, by combining \autoref{eq:nonsecasTDnorm} (with a judicious sign change) with  \autoref{eq:BasFT}, and defining $t_\pm = \frac{t_1\pm t_2}{2}$, $\nu_\pm = \nu_1\pm \nu_2$, that
\begin{align}
\norm{B - B_{\rm sec}}_\infty &\leq  \left| \int_{-\infty}^\infty dt_1 \int_{-\infty}^\infty dt_2\, \mathcal{W} (t_1, t_2) \right| 
\nonumber\\ &=\left| \int_{-\infty}^\infty dt_1 \int_{-\infty}^\infty dt_2\, \mathcal{W} (-t_1, t_2) \right|
\nonumber\\ &= \left| \int_{-\infty}^\infty dt_1 \int_{-\infty}^\infty dt_2 \int_{-\infty}^\infty \frac{d\nu_1}{2\pi}  \int_{-\infty}^\infty \frac{d\nu_2}{2\pi} b_+({\nu_+}) b_-({\nu_-})\left( 1- w\left(\frac{\nu_-}{\mu}\right)\right) e^{i\nu_1 t_1 + i\nu_2 t_2}\right|
\nonumber\\&\leq  \int_{-\infty}^\infty dt_+ \left| \int_{-\infty}^\infty \frac{d\nu_+}{2\pi} b_+(\nu_+) e^{i \nu_+ t_+}\right|\times
\int_{-\infty}^\infty dt_- \left| \int_{-\infty}^\infty  \frac{d\nu_-}{2\pi} b_-({\nu_-})\left( 1- w\left(\frac{\nu_-}{\mu}\right)\right) e^{i\nu_-t_-}\right|
\nonumber\\&\leq \int_{-\infty}^\infty dt_+ |\widehat{b_+}(t_+)| \times \int_{-\infty}^\infty dt_- \left| \int_{-\infty}^\infty  \frac{d\nu_-}{2\pi} b_-({\nu_-})\left( 1- w\left(\frac{\nu_-}{\mu}\right)\right) e^{i\nu_-t_-}\right|\label{eq:workingsecular}
\end{align}
where $\widehat{b_+}(t_+)$ is the inverse Fourier transform of $b_+(\nu_+)$. 

We find that in the two cases of interest to us, $|\widehat{b_+}(t_+)|$ takes the simple form
\begin{equation}\label{eq:app:gplus_explicit}
|\widehat{b_+}(t_+)| = |\widehat{h_+}(t_+)| = |\widehat{g_+}(t_+)| =  \frac{1}{\sqrt{\pi}\sigma}  e^{\beta^2/4\sigma^2} e^{-4t_+^2/\sigma^2}
\end{equation}
Therefore, we can express the first term in the product on the RHS of the final inequality \autoref{eq:workingsecular} above as
\begin{equation}
\int_{-\infty}^\infty dt_+ |\widehat{b_+}(t_+)|  = \frac{e^{\beta^2/4\sigma^2}}{2}\label{eq:nuplusbound}
\end{equation}
in both the coherent and exact cases, which will only give an overall constant prefactor.

Perhaps unsurprisingly, the nontrivial bound involves the integration over $t_-$ in \autoref{eq:workingsecular}. It is useful to split this into a short-time contribution $\mathcal{I}_1$ with $t\in [-t_0, t_0]$ and a long-time contribution $\mathcal{I}_2$ with  $|t|>t_0$, with $t_0$ to be specified below. The short-time piece is given by 
\begin{align}
     \mathcal{I}_1 \equiv \int_{-t_0}^{t_0} dt_- \left| \int_{-\infty}^\infty  \frac{d\nu_-}{2\pi} b_-({\nu_-})\left( 1- w\left(\frac{\nu_-}{\mu}\right)\right) e^{i\nu_-t_-}\right| &\leq 2t_0  \int_{-\infty}^\infty  \frac{d\nu_-}{2\pi} \left| b_-({\nu_-})\right|\left| 1- w\left(\frac{\nu_-}{\mu}\right)\right|.\label{eq:shorttime1}
\end{align}
Given our choice of bump function, $1- w\left(\frac{\nu_-}{\mu}\right)$ is only nonzero for $\nu_-\gtrsim \mu$, where it is close to $1$, so that the integral is dominated by the tails, and we have
\begin{equation}
\int_{-\infty}^\infty  \frac{d\nu_-}{2\pi} \left| b_-({\nu_-})\right|\left| 1- w\left(\frac{\nu_-}{\mu}\right)\right|  = O\left(2 \int_{\mu}^\infty \frac{d\nu_-}{2\pi} \left| b_-({\nu_-})\right|\right) = O\left(\frac{e^{-\frac{\sigma^2\mu^2}{16}}}{\sigma^2 \mu} \right)\label{eq:shorttime2}
\end{equation}
where in the second step  we have used the fact that both $|b_-|\leq e^{-(\sigma \nu_-)^2/16}$ for either choice $b_- = g_-, h_-$. Combining \autoref{eq:shorttime1} and \autoref{eq:shorttime2}, we have
\begin{equation}
    \mathcal{I}_1 = O\left(\frac{t_0 e^{-\frac{\sigma^2\mu^2}{16}}}{\sigma^2 \mu} \right)\label{eq:I1bound}
\end{equation}

We turn next to the long-time contributions. By integrating by parts twice, we observe that
\begin{align}
\int_{-\infty}^\infty  \frac{d\nu_-}{2\pi} b_-({\nu_-})\left( 1- w\left(\frac{\nu_-}{\mu}\right)\right) e^{i\nu_-t_-} = -\frac{1}{t_-^2}\int_{-\infty}^\infty  \frac{d\nu_-}{2\pi}\, \frac{d^2}{d\nu_-^2} \left[b_-({\nu_-})\left( 1- w\left(\frac{\nu_-}{\mu}\right)\right)\right] e^{i\nu_-t_-}.
\end{align}
We can then use this result to obtain a useful bound on the long-time contributions,
\begin{align}
    \mathcal{I}_2 &\equiv \int_{\mathbb{R}\backslash[-t_0,t_0]}dt_-\left|\int_{-\infty}^\infty  \frac{d\nu_-}{2\pi} b_-({\nu_-})\left( 1- w\left(\frac{\nu_-}{\mu}\right)\right) e^{i\nu_-t_-}\right| \nonumber\\
    &\leq \frac{2}{t_0} \int_{-\infty}^\infty  \frac{d\nu_-}{2\pi}\, \left|\frac{d^2}{d\nu_-^2} \left[b_-({\nu_-})\left( 1- w\left(\frac{\nu_-}{\mu}\right)\right)\right] \right|.\nonumber\\ \label{eq:intermediateboundlongtime}
\end{align}
Using the product rule, we have that
\begin{equation}
\frac{d^2}{d\nu_-^2} \left[b_-({\nu_-})\left( 1- w\left(\frac{\nu_-}{\mu}\right)\right)\right] =\left( 1- w\left(\frac{\nu_-}{\mu}\right) \right) b_-''(\nu_-) + \frac{1}{\mu} b'_-(\nu_-) w'\left(\frac{\nu_-}{\mu}\right) + \frac{1}{\mu^2} b_-(\nu_-) w''\left(\frac{\nu_-}{\mu}\right), \label{eq:productrule}
\end{equation}
so that on inserting \autoref{eq:productrule} into \autoref{eq:intermediateboundlongtime} and using the triangle inequality that
\begin{align}
 \mathcal{I}_2 &\leq \frac{2}{t_0} \int_{-\infty}^\infty  \frac{d\nu_-}{2\pi}\,\left( \left| \left( 1- w\left(\frac{\nu_-}{\mu}\right) \right) b_-''(\nu_-) \right|
 + \left|\frac{1}{\mu} b'_-(\nu_-) w'\left(\frac{\nu_-}{\mu}\right)  \right| +  \left|\frac{1}{\mu} b_-(\nu_-) w''\left(\frac{\nu_-}{\mu}\right)  \right|\right)\label{eq:productrulesplitbound}
\end{align}

Let us now consider the two choices $b_- =h_-, g_-$ corresponding to the local driving and exact samplers. We have that $h_-(\nu_-) \propto e^{-(\sigma\nu_-)^2/16}$, whereas $h_-(\nu_-) \propto \tanh\left(\frac{\beta\nu_-}4\right) e^{-(\sigma\nu_-)^2/16}$. Crucially, the $\tanh$ that appears in the 
latter case has a bounded derivative everywhere, and so is relatively innocuous when inserted into \autoref{eq:productrulesplitbound}. Therefore,  after a  scaling analysis (that we do not reproduce here as it is straightforward but tedious) we can show that each term in \autoref{eq:productrulesplitbound} takes the form ${\rm Lpoly}(\beta\nu, \sigma\nu) \times e^{-(\sigma\nu_-)^2/16}$, where ${\rm Lpoly}$ represents some Laurent polynomial of bounded negative and positive degrees. Meanwhile, for our choice of bump function, $w'(\nu_-/\mu)$ and $w''(\nu_-/\mu)$ are both O(1) around $\nu_-/\mu\sim 1$ or vanish otherwise, whereas as stated before $1-w(\nu_-/\mu)$ is only nonzero outside the bounded domain $[-\mu, \mu]$. 
Combining these results, it is evident that we can bound the various terms in \autoref{eq:productrulesplitbound} as
\begin{equation}
\mathcal{I}_2 = O\left(\frac{\sigma}{t_0} {\rm Lpoly}(\beta\mu, \sigma\mu)e^{-(\sigma\mu)^2/16}\right) \label{eq:I2bound}
\end{equation}
which results from considering that the integrals only receive contributions either near $\pm \mu$ or $(\pm \mu, \infty)$, rescaling the integrand, and estimating the weight due to the Gaussian factors, and the $\beta$-dependence will only appear for the exact sampler.  Combining the bound \autoref{eq:nuplusbound} on the $\nu_+$ integral  with the bounds \autoref{eq:I1bound} and \autoref{eq:I2bound} respectively and setting $\sigma = t_0$
\begin{equation}
    \norm{B - B_{\rm sec}}_\infty = O\left( {\rm Lpoly}(\beta\mu, \sigma\mu)e^{-(\sigma\mu)^2/16}\right)
\end{equation}

Again, for the case of multiple jump operators $\nB = |\mathcal{A}|$ the errors add independently, so that we have
\begin{equation}\label{eq:bound_non_sec}
    \norm{B - B_{\rm sec}}_\infty = O\left( \nB\,{\rm Lpoly}(\beta\mu, \sigma\mu)e^{-(\sigma\mu)^2/16}\right)
\end{equation}

Now, substituting \autoref{eq:bound_non_sec} and \autoref{eq:bound_almost_commuting} into \autoref{eq:errorintermsofsecnorms} we obtain
\begin{equation}
    \norm{\rho_{\infty} - \rho_{\beta}}_1 = O \left( 
    \nB\, \left[
        \beta \mu + {\rm Lpoly}(\beta\mu, \sigma\mu)e^{-(\sigma\mu)^2/16}
    \right] \max(t_{\rm mix}(\mathcal{L}_\infty), t_{\rm mix}(\mathcal{L}_{\rm diss})
    \right)
\end{equation}
which for some choice of $\mu = \Tilde O(\sigma^{-1})$ (the $\Tilde O$ now hiding extra-log factors necessary to suppress the second term above) yields the desired result in \autoref{eq:app:bound_secular_final}.

\subsubsection{Choice of Filter Functions and Kernels.} 

To complete the proof, we must obtain the relevant functions $b_+, b_-$ for the two samplers in question, for a specific filter function, and confirm that they satisfy the properties assumed in the proof above. While this can be done for arbitrary sufficiently quickly decaying $f(t)$, for concreteness here we present results for the specific choice \autoref{eq:chosenfilterfunction} in the main text.

With this choice, we see that the Lamb shift kernel in \autoref{eq:LSkernel} is given by (taking $T\to\infty$ as we have throughout the secular approximation, and dropping the superscript $T$)
\begin{align}
-2i h_{\nu_1, \nu_2} &= \int_{-{\infty}}^{\infty}d t_1 \int_{-{\infty}}^{\infty}d t_2\, \text{sgn}(t_1-t_2) f(t_1) f^*(t_2)  e^{i\nu_1 t_1 - i \nu_2 t_2} \nonumber\\
&= \int_{-\infty}^\infty \mathrm{d}t_1\,\int_{-\infty}^\infty \mathrm{d}t_2\, {\rm sgn}(t_1+t_2)f(t_1)f^*(-t_2)e^{i \nu_1 t_1} e^{i \nu_2 t_2}\nonumber\\
	&= 2 \int_{-\infty}^\infty \mathrm{d}t_+\,\int_{-\infty}^\infty \mathrm{d}t_-\, {\rm sgn}(t_+)f(t_++t_-)f^*(-t_++t_-) e^{i \nu_+ t_+} e^{i \nu_- t_-}
\end{align}
We can further simplify the integrand as follows:
\begin{align}
	f(t_++t_-)f^*(-t_++t_-)&=\frac{2}{\pi \sigma^2}\exp\left\{-\frac{2}{\sigma^2}\left[\left(\left(t_+-\frac{i\beta}{4}\right)+t_-\right)^2+\left(\left(t_+ - \frac{i\beta}{4}\right)-t_-\right)^2\right]\right\}\nonumber\\
    &
	=\frac{2}{\pi \sigma^2}\exp\left[-\frac{4}{\sigma^2}\left(t_+-\frac{i\beta}{4}\right)^2\right]
\exp\left[-\frac{4}{\sigma^2}t_-^2\right].
\end{align}
From this, we obtain 
\begin{align}
\begin{split}
	h_{\nu_1,\nu_2}=
	h_+(\nu_+)h_-(\nu_-).
\end{split}
\end{align}
Here,
\begin{align}
	h_+(\nu_+)=-\frac{1}{2 i}\int_{-\infty}^\infty \mathrm{d}t_+ \frac{2}{\sqrt{\pi}  \sigma}{\rm sgn}(t_+)\exp\left[-\frac{4}{\sigma^2}\left(t_+ -\frac{i \beta}{4}\right)^2\right]
	e^{-i\nu_+t_+}
\end{align}
with gives us that 
\begin{equation}
|\widehat{b_+}(t_+)| = \left|\frac{i}{\sqrt{\pi} \sigma}{\rm sgn}(t_+)\exp\left[-\frac{4}{\sigma^2}\left(t_+ - \frac{i \beta}{4}\right)^2\right]\right| =\frac{1}{\sqrt{\pi}\sigma}  e^{\beta^2/4\sigma^2} e^{-4t_+^2/\sigma^2}
\end{equation}
and hence satisfies the assumption of Gaussian decay at large $t_+$. Meanwhile,  we have
\begin{align}
h_-(\nu_-)=\int_{-\infty}^\infty \frac{\mathrm{d}t_-}{\sqrt{\pi} \sigma} e^{-\frac{4}{\sigma^2}t_-^2} e^{-i\nu_-t_-}=\frac{1}{2} e^{-\frac{1}{16}\sigma^2\nu_-^2}.
\end{align}
which has the form of a bounded function (in this case a constant) multiplied by a Gaussian.

For the exact sampler, instead, we have that $\alpha_{\nu_1, \nu_2} = \exp[ -(\sigma\nu_1)^2/8 -(\sigma\nu_2)^2/8 -\beta\nu_1/4-\beta\nu_2/4]$, which yields
\begin{equation}
g_{\nu_1,\nu_2} =  -\frac{1}{2i} \tanh\frac{\beta(\nu_1-\nu_2)}{4}\alpha_{\nu_2, \nu_2} = g_+(\nu_+) g_-(\nu_-)
\end{equation}
with
\begin{equation}
g_+(\nu_+) = \frac{i}{2} e^{-\beta\nu_+/4} e^{-\frac{(\sigma \nu_+)^2}{16}}\quad \text{and} \quad g_-(\nu_-) = \tanh \frac{\beta\nu_-}{4} e^{- \frac{(\sigma\nu_-)^2}{16}}
\end{equation}
Evidently, $g_-$ once again has the form of a bounded function multiplying a Gaussian. Turning finally to the Fourier transform fo $g_+$, we have
\begin{align}
|\widehat{g_+}(t_+)| &= \left| \frac{i}{2}\int_{-\infty}^\infty \frac{d\nu_+}{2\pi}e^{-\beta\nu_+/4} e^{-\frac{(\sigma \nu_+)^2}{16}} e^{i\nu_+t_+}\right| \nonumber\\&= \left| \frac{i}{\sqrt{\pi}\sigma}\exp\left[-\frac{4}{\sigma^2}\left(t_+ - \frac{i \beta}{4}\right)^2\right] \right| \nonumber\\&= \frac{1}{\sqrt{\pi} \sigma} e^{\beta^2/4\sigma^2} e^{-4t_+^2/\sigma^2}
\end{align}
which is identical to $|\widehat{h_+}(t_+)|$ and in particular of exactly the form assumed in \autoref{eq:app:gplus_explicit}.

\section{Bounding the error in dropping the ``rewinding''\label{app:rewindingerror}}

In this appendix, we show that dropping the rewinding procedure in our protocol adds to the total error bound in a controllable way, namely that the error bound for the original protocol increases by a factor of $2\tmix$ [cf. \autoref{eq:rewindingerror} in the main text].

Concretely, we consider a channel $\mathcal K'$ defined as
$\mathcal K'[\sigma] \equiv V \mathcal K[\sigma] V^{\dagger}$ for some unitary $V$, and a state $\rho_{\beta}$ that invariant under conjugation with $V$: $V \rho_{\beta} V^{\dagger} = \rho_\beta$. We assume that fixed point of $\mathcal K$, $\rho$, ($\mathcal K[\rho] = \rho$) is close to $\rho_{\beta}$: $\norm{\rho - \rho_{\beta}}_1 \leq \delta$ for some $\delta > 0$.  We then show that
\begin{equation}\label{eq:app:rewinding_final}
    \norm{\rho' - \rho_{\beta}}_1 \leq 4\delta \tmix
\end{equation}
where $\rho'$ is the fixed-point of $\mathcal K'$ and $\tmix$ the mixing time of $\mathcal K$, respectively.

As preliminaries, observe that since the channel $\mathcal K$ is contracting w.r.t. to the Schatten 1-norm,
\begin{equation} 
    \norm{\mathcal K[\rho_{\beta}] - \rho}_1 
    = \norm{\mathcal K[\rho_{\beta}] - \mathcal K[\rho]}_1
    \leq \norm{\rho_{\beta} - \rho}_1 
    \leq \delta.
\end{equation}
Furthermore, because the Schatten 1-norm is preserved under unitary conjugation,
\begin{equation}
    \norm{\mathcal K'[\rho_{\beta}] - \rho_{\beta}}_1
    = \norm{V \mathcal K[\rho_{\beta}] V^\dagger - \rho_{\beta}}_1 
    \leq \norm{V \left(\mathcal K[\rho_\beta] - \rho \right) V^\dagger}_1
        + \norm{V \left(\mathcal \rho - \rho_{\beta} \right) V^\dagger}_1
    \leq 2\delta.
\end{equation}

Now, for the fixed point $\rho'$ of $\mathcal K'$, we have
\begin{equation}\label{eq:app:rewinding_intermediate1}
    \norm{\rho' - \rho_{\beta}}_1 
    = \norm{\mathcal K'[\rho'] - \rho_{\beta}}_1
    \leq \norm{\mathcal K'[\rho'] - \mathcal K'[\rho_{\beta}]}_1
        + \norm{\mathcal K'[\rho_{\beta}] - \rho_{\beta}}_1
    \leq \eta \norm{\rho' - \rho_{\beta}}_1  + 2 \delta
\end{equation}
where $\eta \in [0, 1)$ is the \emph{contraction} factor
\begin{equation}
    \eta' \equiv 
    \sup_{\omega\neq \zeta} \frac{\norm{\mathcal K'[\omega] - \mathcal K'[\zeta]}_1}{\norm{\omega - \zeta}_1} 
    = \sup_{\omega\neq \zeta} \frac{\norm{\mathcal K[\omega] - \mathcal K[\zeta]}_1}{\norm{\omega - \zeta}_1} = \eta
\end{equation}
where the second inequality follows from the definition of $\mathcal K$.
Note that $\eta < 1$ under the assumption that $\mathcal K$ is primitive (that is irreducible and aperiodic), i.e. that any initial state converges to a unique fixed point.
Rearranging \autoref{eq:app:rewinding_intermediate1} then gives
\begin{equation}
    \norm{\rho' - \rho_{\beta}}_1 \leq \frac{2\delta}{1-\eta}.
\end{equation}

What is left to show to obtain \autoref{eq:app:rewinding_final} is then to relate the contraction factor $\eta$ to the mixing time $\tmix$. 
To this end, note that  the contraction factor lower bounds the second-largest absolute value in the spectrum of $\mathcal K$, which in turn lower bounds the mixing time (see e.g. Lemma 30 of Ref. \onlinecite{kastoryano2012quantum_sobolev}):
\begin{equation}
\frac{1}{1-\eta} \leq \frac{1}{1-\abs{\lambda_1}} \leq 2 \tmix
\end{equation}
Substituting this above then yields the final result
\begin{equation}
    \norm{\rho' - \rho_{\beta}}_1 
    \leq \frac{2\delta}{1-\eta}
    \leq 4 \delta \tmix
\end{equation}

\section{Mixing time analysis}\label{subsec:Mixingtimeanalysis}

\begin{figure*}[t!]
    \centering   \includegraphics{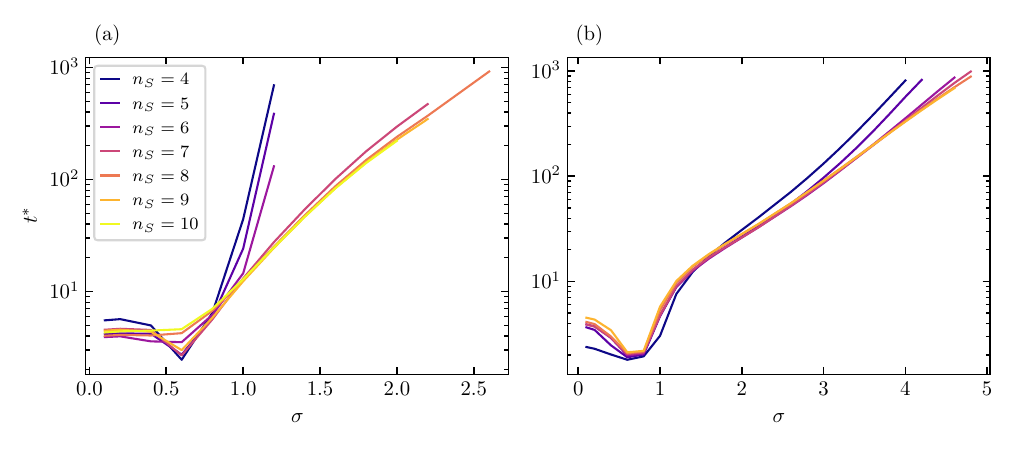}
    \caption{The mixing time of the maximally mixed state $t^*$ as a function of the parameter $\sigma$ in the filter function, $\beta=1.0$. (a) For the mixed-field Ising model defined in Eq.~\eqref{eq:MFI}, (b) the transverse field Ising model defined in Eq.~\eqref{eq:TFI}. For large $\sigma$, $t^*$ scales exponentially with $\sigma$.
}
    \label{fig:Numericsmixingtime}
\end{figure*}
In this section, we present numerical results for the mixing time of the exact Gibbs sampler $\cL_\beta$ as a function of the parameter $\sigma$ in the filter function and the system size $n_S$. The analysis is important, since a different choice of $\sigma$ in our protocol leads to different jump operators in the approximated Lindbladian and as such quantities appearing in the accuracy bounds as the mixing time are not necessarily independent of the mixing time.

Since the mixing time itself is difficult to probe numerically, we use as a proxy the time of convergence, starting from the maximally mixed state $\rho_0$:
\begin{align}
    t^*=\min_t \norm{\rho_\beta-e^{\cL_\beta t}\rho_0}_1<0.1,
\end{align}
with $\rho_0$ being the maximally mixed state. $t^*$ serves thus as a lower bound for the true mixing time.
We compare $t^*$ for the mixed-field Ising models defined in Eq.~\eqref{eq:MFI} and the transverse field Ising model:
\begin{align}\label{eq:TFI}
    H=\sum_i Z_iZ_{i+1}+g X_i
\end{align}
with $g=0.9045$.
While the mixed field Ising model is known to be thermalizing, the transverse field Ising model can be mapped to a model of free fermions, and the energy spectrum is described by discrete excitations with a finite spectral gap for $g\neq1$.
If $\sigma$ is smaller than the spectral gap, we thus expect that transitions of these excitations are suppressed and the mixing time exponentially increases. 
The same argument essentially also holds for the mixed-field Ising model, at small temperatures below the gap. At high or intermediate temperatures, however, the effect of changing the filter is less obvious, since the level spacing at finite energy density is expected to be exponentially small in system size.  

We choose $Y_i$ on each site, as the set of jump operators, and $\beta=1.0$. 
The results are shown in Fig.~\ref{fig:Numericsmixingtime}.
For the transverse-field Ising model, $t^*$ increases exponentially with $\sigma$ with almost no dependence on system size. For the mixed-field Ising model, initially $t^*$ \emph{decreases} with system size, but for $n_S\geq 7$, this decrease saturates and we again observe an exponential increase of $t^*$ with $\sigma$ with little dependence on system size.
Understanding the dependence of the mixing time on parameters, such as the parameter $\sigma$ in the filter function, remains to be investigated in future work.

\end{widetext}

\end{appendix}
\twocolumngrid

\bibliography{references_paper}

\end{document}

%% file: preamble.tex
\usepackage[utf8]{inputenc}
\usepackage{fullpage}
\usepackage{graphicx}
\usepackage{amsmath}
\usepackage{amsthm}
\usepackage{enumitem}
\usepackage{amssymb}
\usepackage{wasysym}
\usepackage{oplotsymbl}
\usepackage{bm} % bold italic vectors
\usepackage{dsfont} % identity operator
\usepackage{mathrsfs} % mathscr font
\usepackage{physics}
\usepackage{mathtools}
\usepackage{quantikz}

\usepackage[dvipsnames]
{xcolor}
\definecolor{darkblue}{rgb}{0,0,.65}
\definecolor{darkgreen}{rgb}{0.28,0.41,0.19}
\definecolor{nicegreen}{rgb}{0.28,0.85,0.19}

\usepackage[colorlinks=true,linkcolor=blue,allcolors=blue]{hyperref}
\usepackage[capitalize,nameinlink]{cleveref}

\usepackage{comment}
\usepackage[normalem]{ulem}

\def\equationautorefname~#1\null{Eq. (#1)\null}

\newcommand{\appref}[1]{\hyperref[#1]{App.~\ref*{#1}}}

\newcommand{\tmix}{\tau_{\rm mix}}% mixing time
\newcommand{\tmixmax}{\tau^*_{\rm mix}}% maximum of all mixing times involved
\newcommand{\bohr}{{\mathfrak B}}
\newcommand{\nB}{n_{\rm B}}
\newcommand{\nS}{n_{\rm S}}

\newcommand{\e}{\mathrm{e}}
\newcommand{\I}{\mathrm{i}}
\newcommand{\rd}{\mathrm{d}}
\newcommand{\cL}{\mathcal{L}}
\newtheorem{prototheorem}{Theorem}[section]

\newtheorem{protolemma}[prototheorem]{Lemma}